\documentclass[floatfix,amsmath,amssymb,aps,prc,twocolumn,superscriptaddress,showpacs,preprintnumbers,nofootinbib]{revtex4-1}

\usepackage{graphicx,color}
\usepackage{multirow}
\usepackage{times}
\usepackage{bm}
\usepackage{epstopdf}
\usepackage{enumerate}
\usepackage{ulem}

\newcommand{\comment}[1]{}

\renewcommand\sout{\bgroup \color{red} \ULdepth=-.5ex \ULset}

\begin{document}

\title{
Mean-field update in the JAM microscopic transport model:
Mean-field effects on collective flow
in high-energy heavy-ion collisions at $\sqrt{s_{NN}}=2-20$ GeV energies
}
\author{Yasushi Nara}
\affiliation{
Akita International University, Yuwa, Akita-city 010-1292, Japan}

\author{Akira Ohnishi}
\affiliation{Yukawa Institute for Theoretical Physics, Kyoto University,
Kyoto 606-8502, Japan}

\date{\today}

\preprint{YITP-21-97}
\begin{abstract}
The beam energy dependence of the directed flow is a sensitive probe for the properties of strongly interacting matter.
Hybrid models that simulate dense (core) and dilute (corona) parts of the system
by combining the fluid dynamics and hadronic cascade model describe well
the bulk observables, strangeness productions, and radial and elliptic flows
in high-energy heavy-ion collisions in the high baryon region.
However, the beam energy dependence of the directed flow cannot be described in existing hybrid models.
We focus on improving the corona part, i.e., the nonequilibrium evolution part of the system, by introducing the mean-field potentials into a hadronic cascade model.
For this purpose,
we consider different implementations of momentum-dependent hadronic mean fields
in the relativistic quantum molecular dynamics (RQMD) framework.
First, Lorentz scalar implementation of a Skyrme type potential is examined. 
Then, full implementation of the Skyrme type potential as a Lorentz vector
in the RQMD approach is proposed.
We find that scalar implementation of the Skyrme force is too weak
to generate repulsion explaining observed data of sideward flows at $\sqrt{s_{NN}}<10$ GeV, 
while vector implementation gives collective flows compatible with the data for a wide range of beam energies $2.7 <\sqrt{s_{NN}}<20$ GeV.
We show that our approach reproduces
the negative proton directed flow at $\sqrt{s_{NN}}>10$ GeV discovered
by experiments.
We discuss the dynamical generation mechanisms of the directed flow
within a conventional hadronic mean field.
A positive slope of proton directed flow is generated predominantly 
during compression stages of heavy-ion collisions
by the strong repulsive interaction due to high baryon densities.
In contrast, in the expansion stages of the collision,
the negative directed flow is generated more strongly than the positive one
by the tilted expansion and shadowing by the spectator matter.
At lower collision energies $\sqrt{s_{NN}}<10$ GeV, the positive flow wins
against the negative flow because of a long compression time.
On the other hand, at higher energies $\sqrt{s_{NN}}>10$ GeV, 
negative flow wins because of shorter compression time and longer
expansion time. A transition beam energy from positive to negative flow
is highly sensitive to the strength of the interaction.
\end{abstract}

\maketitle

\section{Introduction}

The phase structure of QCD matter
for a wide range of baryon density
is of fundamental interest~\cite{cbmbook}.
Understanding the equation of state (EoS) for QCD matter is a primary goal.
In particular, a first-order phase transition and a critical point
at finite baryon densities
are predicted by several effective models ~\cite{DHRischke2004}.
The properties of QCD matter have been explored experimentally
using high energy nuclear collisions under various conditions:
beam energy, centrality, and system size dependence.
Currently, high-energy heavy-ion experiments represent one of the most active areas:
heavy-ion experiments are being performed from a few GeV to TeV beam energies
in the same era.
We now have a vast body of data
including different centralities and system sizes from many experiments 
by the accelerators such as the SIS~\cite{SIS} heavy-ion synchrotron the Alternating Gradient Synchrotron (AGS)~\cite{AGS}, the Super Proton Synchrotron (SPS)~\cite{SPS},
the Relativisitic Heavy Ion Collider (RHIC)~\cite{RHIC},
the Large Hadron Collider (LHC)~\cite{LHC}, and so on.
Future facilities such as
the Facility for Antiproton and Ion Research (FAIR)~\cite{FAIR}, the Nuclotron-based Ion Collider fAcility (NICA)~\cite{NICA},
the High Intensity Heavy-ion Accelerator Facility
(HIAF)~\cite{Yang:2013yeb},
and the J-PARC Heavy Ion Project (J-PARC-HI)~\cite{HSakoNPA2016}
are being constructed or are planned to perform high precision measurements.

Anisotropic collective flows are considered to be a good probe to extract
the EoS of dense QCD matter~\cite{Stoecker:1986ci,Ollitrault:1992,Danielewicz:1998vz,Danielewicz:2002pu,Stoecker:2004qu}.
Noncentral collisions create azimuthally asymmetric excited matter,
and subsequent collective expansion results in azimuthally anisotropic emission of particles.
The distribution can be analyzed from the coefficients 
in the Fourier expansion of measured particle spectra~\cite{Poskanzer:1998yz}.
The directed flow is defined by the first coefficient $v_1=\langle\cos\phi\rangle$,
and the second coefficient $v_2=\langle\cos2\phi\rangle$ is called the elliptic flow,
where $\phi$ is the azimuthal angle of an outgoing particle with respect to
the reaction plane.
These flows have been measured by various experiments, and now we have excitation
functions of flows from $\sqrt{s_{NN}}\approx2$ GeV to 5 TeV.
The proton elliptic flow is negative 
below $\sqrt{s_{NN}}\approx3$ GeV due to a shadowing of spectator matter (squeezed out)
and it becomes positive at higher beam energies~\cite{Danielewicz:1998vz,E895v2}.
Transport theoretical models describe this sign change of
the elliptic flow~\cite{Danielewicz:1998vz,E895v2}.
On the other hand, the data show that the slope of the proton $v_1(y)$
with respect to rapidity $y$ is positive (normal or positive flow)
$dv_1/dy>0$ up to the beam energy of
$\sqrt{s_{NN}}\approx10$ GeV, and then it becomes negative
(antiflow or negative flow) 
above 10 GeV at mid-rapidity~\cite{NA49prc,STARv1,Singha:2016mna}.

It has been argued that the negative directed flow could be an effect of the
softening of the EoS, and it may be a signature of a first-order-phase transition%
~\cite{Rischke:1995pe,Brachmann:1999xt,Csernai1999}.
Fluid dynamical simulations and microscopic transport models
predict that the softening happens at around a
beam energy less than 5 GeV%
~\cite{Rischke:1995pe,Brachmann:1999xt,Csernai1999,Li:1998ze,Nara:2016hbg},
which is inconsistent with the experimental data.
On the other hand, antiflow at beam energies above 27 GeV is naturally explained by the
transport models~\cite{Konchakovski:2014gda}
by the combination of space-momentum correlations together with the correlation
between the position of a nucleon in the nucleus and its stopping~\cite{Snellings:1999bt}.
The direct reason for the negative slope is the tilted matter created
in noncentral collisions, which generates antiflow predominately over the normal flow during the expansion stage.
The color glass condensate model also predicts twisted matter~\cite{Adil:2005qn}.
The tilted source was used in the initial condition of the hydrodynamical
evolution to explain negative directed flow at the top RHIC energy~\cite{Bozek:2010bi}.
The transport models describe the directed flow below 7.7 GeV  or above 27
GeV~\cite{Konchakovski:2014gda,Nara:2019qfd,Nara:2020ztb}.
The three-fluid dynamics (3FD) model~\cite{Ivanov:2014ioa} reproduces the rapidity dependence of the
directed flow at 11.5 GeV with the crossover and first-order phase transition
scenario. A transport calculation with attractive trajectory prescription
\cite{Nara:2016phs} also fits the data at 11.5 GeV.
However, none of the fluid models explain the beam energy dependence of the slope so far.

An alternative way to understand
the space-time evolution of the matter created 
in high-energy nuclear collisions is to utilize
microscopic transport models such as Boltzmann-Uehling-Uhlenbeck (BUU)~\cite{Bertsch:1984gb,GiBUU}
and quantum molecular dynamics (QMD)~\cite{Aichelin:1986wa,Aichelin:2019tnk} approaches
and their relativistic versions, RBUU~\cite{RBUU,RVUU} and
RQMD~\cite{RQMD1989,RQMDmaru,Fuchs:1996uv,Maruyama:1996rn,Isse:2005nk,Nara:2019qfd,Nara:2020ztb}, 
which have been developed and successfully employed to understand the nonequilibrium collision dynamics of
high-energy nuclear collisions.
The main two ingredients of the microscopic transport model are the Boltzmann type collision term 
and the mean-field interaction.
Later, hybrid models were developed by combining fluid dynamics
into a microscopic transport model (in the cascade mode)%
~\cite{Petersen:2008dd,Steinheimer:2014pfa,Karpenko:2015xea,%
Batyuk:2016qmb,Denicol:2018wdp,Akamatsu:2018olk,Shen:2020jwv}
to describe heavy-ion collisions in high baryon density regions.
The inclusion of fluid dynamics improves the particle multiplicities, especially strangeness particle and antibaryon yields, and reproduces the beam energy dependence of the elliptic flow by changing the shear viscosity
~\cite{Karpenko:2015xea,Shen:2020jwv}. Thus, hybrid models describe well the radial and elliptic flows.
However, hybrid models do not reproduce the beam energy dependence of the directed flow~\cite{Steinheimer:2014pfa,Shen:2020jwv}.

We have developed a dynamically integrated hybrid JAM + hydro model~\cite{Akamatsu:2018olk} by utilizing the dynamical initialization of the fluid
~\cite{Okai:2017ofp,Kanakubo:2018vkl,Kanakubo:2019ogh,Kanakubo:2021qcw},
which shows the importance of the separation of the dense and dilute part (core-corona separation) of the system. In other words, the nonequilibrium evolution of the system exists for all stages of the collision at the collision energy of
the high baryon density region. Recently, it was found in Ref.~\cite{Kanakubo:2021qcw}
that this is also true even for the LHC energies.
However, all hybrid models use cascade models for the nonequilibrium evolution, in which EoS is an ideal gas of hadrons up to small corrections from strings.
Thus, the corona part described by the particles does not have the correct EoS effects.
It is well known that the cascade model lacks the magnitude of pressure, and mean-field effects
are necessary to reproduce the experimentally observed anisotropic flows~\cite{Danielewicz:1998vz,Danielewicz:2002pu}.

Thus none of the existing dynamical models including fluid, hadronic cascade, 
hybrid, and integrated models explain the beam energy dependence of the slope of the directed flow so far.
Now the question arises, what is the reason for the transition from
positive to negative slope at around 10 GeV?
From the discussions above,
the mean field in the nonequilibrium evolution in a hybrid model
should improve the description of the collision dynamics
in the high baryon density regions. 
Especially, the mean field in the nonequilibrium compression stages of the collision
may significantly alter the dynamics.
Another important aspect is the formation of tilted matter, which induces
the negative flow in the expansion stage.
The balance of the positive flow from large pressure in the compression stage 
and the negative flow from the formed tilted matter may cause the nonmonotonic
beam energy dependence of the directed flow slope.

It should be noted that 
the origin of the nonmonotonic behavior of the directed flow slope is relevant
to the onset energy,
at which the partonic degrees of freedom becomes significant
during heavy-ion collisions.
The fraction of partonic matter is expected to increase gradually
with increasing beam energy, since the system volume is finite in heavy-ion collisions and a sharp phase transition will not take place even if it exists
in the thermodynamic limit.
In hybrid models (the 3FD model), 
a large part of the fluid (participant fluid) represents
the partonic matter,
while the cascade part (the spectator fluid)
describes hadronic matter.
Since a part of the core can be hadronic, the dominance of the core part
or the negative flow may be regarded as the lower bound of the onset energy.
We consider that a significant part of matter would still be hadronic at the colliding
energies studied in this article, $\sqrt{s_{NN}}=2.7-20~\mathrm{GeV}$,
while partonic degrees of freedom can appear to some extent.

In this work, we shall concentrate on the nonequilibrium evolution of the system by employing a newly developed microscopic transport model, JAM2.
We employ an RQMD approach to take account of the mean-field effects.
It is well known that collective flows are highly sensitive to
the mean-field interactions at high baryon density regions~\cite{Danielewicz:2002pu,Sorge1997}.
The RQMD model is an $N$-body theory, which describes the multihadron
interactions, which are realized by the 
interactions among $N$ particles.
Boltzmann-type collision terms are also incorporated in RQMD.

We extend our version of the RQMD (RQMD/S) model,
which was developed by incorporating momentum-dependent potential~\cite{Isse:2005nk}
into the RQMD/S model~\cite{Maruyama:1996rn}.
In Ref.~\cite{Isse:2005nk},
we showed that the RQMD/S model describes both directed and elliptic flow
for a wide range of beam energies, emphasizing the importance of the
momentum-dependent potential.
However, we found a mistake that the density-dependent part of the force was
overestimated by a factor of 2.
After correcting this mistake, it turns out that the flow from the RQMD/S
model is not as large as the experimental data.
Thus, one of the purposes of this paper is to update our previous results.
Then, we shall propose a more consistent implementation of the Skyrme potential as a Lorentz scalar into the RQMD approach, which we call RQMDs.
The main differences between RQMD/S and RQMDs are the following:
RQMD/S uses an EoS which is obtained by a fully nonrelativistic treatment of the potential, and potentials are implemented in the RQMD framework
under some assumptions to simplify the model.
We correct these defects in RQMDs: it uses EoS from the scalar potential,
and it is consistently implemented into the RQMD framework.
We found that the RQMDs model predictions agree with the RQMD/S results.

Another goal of this paper is to develop 
further a new RQMD approach, in which the Skyme potential is
treated as a fully Lorentz vector (RQMDv).
We shall show that RQMDs generally predicts less pressure required to
explain the experimental flow data, while RQMDv generates stronger pressure
and describes well the beam energy dependence of anisotropic flows.
We will also examine in detail the collision dynamics to understand 
the generation mechanisms of anti-flow at mid-rapidity. 

This paper is organized as follows.
In Sec.~\ref{sec:eos}, we first explain our EoS for three different treatments: the nonrelativistic potential which is used by RQMD/S, the scalar potential for RQMDs, and the vector potential for RQMDv.
In Sec.~\ref{sec:rqmd}, we present how to implement the above constructed EoS into a framework of the RQMD approach.
Then, in Sec.~\ref{sec:results}, we compare the directed and elliptic flows from our models with the experimental data and discuss the generation mechanisms of the directed flow in our model.
Section~\ref{sec:bulk_spectra} is devoted to the study of mean-field effects on the bulk observables.
The summary is given in Sec.~\ref{sec:conclusion}.

\section{Equation of state}
\label{sec:eos}

We start with a short review of equation of state, which
is used in our previous model~\cite{Isse:2005nk},
based on the so-called simplified relativistic quantum molecular dynamics (RQMD/S)
approach.
Then, we present the EoS using Lorentz scalar and vector potentials,
which will be used in the new version of RQMD.

\subsection{Nonrelativistic potential}

We use a Skyrme type density-dependent potential
together with the momentum-dependent potential.
The single-particle potential is given by
\begin{equation}
 U(\rho,p) = U_\mathrm{sk}(\rho) + U_m(p)\,.
\end{equation}
The baryon-density $\rho$ dependent part $U_\mathrm{sk}(\rho)$ is 
assumed to have the following density dependence
\begin{equation}
 U_\mathrm{sk}(\rho) = \alpha\left(\frac{\rho}{\rho_0}\right)
   +\beta\left(\frac{\rho}{\rho_0}\right)^\gamma\,,
   \label{eq:skyrme}
\end{equation}
where the normal nuclear density is taken to be $\rho_0=0.168$ fm$^{-3}$.
The momentum-dependent part $U_m(p)$
is assumed to be given as the momentum folding with the Lorentzian form factor
\begin{equation}
 U_m(\bm{p}) = \frac{C}{\rho_0}\int d^3p' \frac{f(\bm{x},\bm{p}')}{1+[(\bm{p}-\bm{p}')/\mu]^2}
 \label{eq:momdep}
\end{equation}
where $f(x,p)$ is the single-particle distribution function for a nucleon.
In the case of symmetric nuclear matter at zero temperature, it is given by
\begin{equation}
 f(x,p) = \frac{g_N}{(2\pi)^3}\theta(p_f - |\bm{p}|)
\end{equation}
with $g_N=4$ being the degeneracy factor for spin and isospin of nucleons,
and $p_f = \left(\frac{6\pi^2\rho}{g_N}\right)^{1/3}$ is a Fermi momentum.
The energy density at zero temperature is obtained as~\cite{Welke:1988zz}
\begin{align}
 e  &= e_\mathrm{kin} +e_\mathrm{pot},\\
 e_\mathrm{kin} &= \frac{g_N}{(2\pi)^3}\int_0^{p_f} d^3p \sqrt{m_N^2 + p^2}
   \nonumber\\
  &=\frac{g_N}{16\pi^2}\left [
  2p_f^3e_f + m_N^2 e_f p_f - m_N^4\ln\left(\frac{e_f+p_f}{m_N}\right)
  \right] \nonumber\\
  &\simeq \left[\frac{3}{5}\frac{p_f^2}{2m_N} + m_N\right]\rho,
  \\
e_\mathrm{pot}&=   
   \int^\rho_0 U_\mathrm{sk}(\rho')d\rho'
   + \frac{1}{2}\int d^3p U_{m}(p) f(x,p)
   \nonumber\\
   &=\frac{\alpha\rho^2}{2\rho_0}+\frac{\beta\rho^{\gamma+1}}{(\gamma+1)\rho_0^\gamma}
   +\frac{C\rho^2}{2\rho_0}\,F_m\left(\frac{2p_f}{\mu}\right),
   \\
F_m(x)&=\!\frac{6}{x^2}\!\left[\frac{3}{2}\!-\!\frac{4\mathrm{arctan}\,x}{x}\!-\!\frac{1}{x^2}%
\!+\!\frac{(3x^2\!+\!1)\ln (1\!+\!x^2)}{x^4}
\right]\,,
\end{align}
with $m_N$ being a nucleon mass and $e_f=\sqrt{m_N^2+p_f^2}$.
The total potential energy $V$ with density $\rho(\bm{r})$
and phase space $f(r,p)$  distributions is given by
\begin{align}
 V 
 &= \int d^3r \left(\frac{\alpha}{2\rho_0}\rho^2 +
 \frac{\beta}{(\gamma+1)\rho_0^\gamma}\rho^{\gamma+1}
  \right) \nonumber \\
  &+ \frac{C}{2\rho_0}\int d^3rd^3pd^3p' \frac{f(r,p)f(r,p')}{1+[(\bm{p}-\bm{p}')/\mu]^2}
  \label{eq:ematter}
  \,.
\end{align}
It should be noted that the single particle potential (or the real part of the momentum-dependent optical potential) is related to the total potential energy $V$ via the relation
\begin{align}
U_\mathrm{opt}(\bm{r},\bm{p})=\frac{\delta V}{\delta f(\bm{r},\bm{p})}\,.
\end{align}
The total energy per nucleon from the mass is obtained by
\begin{equation}
 \frac{E}{A}  = \frac{e}{\rho} - m_N\,.
\end{equation}

\begin{table*}
\caption{Parameter sets for the Skyrme type potential
in the nonrelativistic, scalar, and vector implementations.
The range parameters in the momentum-dependent part are taken to be
$\mu_1=2.02~\mathrm{fm}^{-1}$ and $\mu_2=1.0~\mathrm{fm}^{-1}$
for MH2 and MS2, independent of the implementation scheme.
For MH1 and MS1, we adopt
$\mu_1=3.173~\mathrm{fm}^{-1}$ (nonrelativistic),
$\mu_1=5.18~\mathrm{fm}^{-1}$ (scalar),
and
$\mu_1=3.23~\mathrm{fm}^{-1}$ (vector).
The optical potential is controlled by the two parameters
$p_0$ and $U_\infty$ via the relations
$U_\mathrm{opt}(\rho_0,p=p_0)=0$
and 
$U_\mathrm{opt}(\rho_0,p=1.7~\mathrm{GeV})=U_\infty$.
In the nonrelativistic and vector implementations,
we take
$(p_0,U_\infty)=(0.65~\mathrm{GeV}, 60~\mathrm{MeV})$.
In the scalar implementation, 
we take
$(p_0,U_\infty)=(0.685~\mathrm{GeV}, 60~\mathrm{MeV})$
for MH1 and MS1
and
$(p_0,U_\infty)=(0.685~\mathrm{GeV}, 50~\mathrm{MeV})$
for MH2 and MS2.
}
\begin{tabular}{c|c|ccccc|ccccc|ccccc}\hline\hline
\multirow{3}{*}{Type} &\multirow{2}{*}{$K$} 
 &\multicolumn{5}{c|}{Nonrelativistic}
 &\multicolumn{5}{c|}{Scalar}
 &\multicolumn{5}{c}{Vector}\\
& & $\alpha$ & $\beta$ & $\gamma$ & $C_1$ & $C_2$ 
  & $\alpha$ & $\beta$ & $\gamma$ & $C_1$ & $C_2$ 
  & $\alpha$ & $\beta$ & $\gamma$ & $C_1$ & $C_2$ 
 \\
 & (MeV)    
 & (MeV)  & (MeV) &  & (MeV) & (MeV)
 & (MeV)  & (MeV) &  & (MeV) & (MeV)
 & (MeV)  & (MeV) &  & (MeV) & (MeV)
\\
 \hline
H   &380 & $-125$  & 70.7 & 2.003 & --       & --    
    & $-123.8$ & 68.75  & 2.124 & --       & --     \\
S   &210 & $-311$  & 256  & 1.203 & --       & --    & $-262.1$ & 206.4  & 1.265 & --       & --     \\
MH1 &380 & $38.6$    & 41.7   & 2.280 & $-169.8$ & --    
         & 168.0     & 49.30  & 2.286 & $-295.2$ & --     
         &  38.95    & 41.71  & 2.273 & $-169.8$ & --    
         \\ 
MS1 &210 & $-207$    & 287    & 1.120 & $-169.8$ & --    
         & $-14.16$  & 230.8  & 1.179 & $-295.2$ & --
         & $-233.1$  & 313.7  & 1.109 & $-169.8$ & --    
         \\
%
MH2 &380 & $-6.56$   & 82.2   & 1.723 & $-386.6$ & 343.9 
         & $-147.5$  & 282.3  & 1.309 & $-850.0$ & 1050.9
         &  $-13.12$ & 88.85  & 1.674 & $-399.0$ & 367.3  \\
MS2 &210 & $-315.1$  & 388.4  & 1.113 & $-386.6$ & 343.9 
         & $-1740$   & 1874.7 & 1.035 & $-850.0$ & 1050.9
         & $-515.7$  & 590.6  & 1.071 & $-399.0$ & 367.3
         \\
\hline\hline
\end{tabular}
\label{table:ns}
\end{table*}

We compare the single-particle potential at the normal nuclear density
\begin{equation}
U_\mathrm{opt}(p)= U(\rho_0,p) = \alpha + \beta + U_{m}(p)\,.
\label{eq:opt}
\end{equation}
with the Schr\"odinger-equivalent optical potential from the Dirac phenomenology%
~\cite{Hama:1990vr}.
%
The parameters of the potentials are fixed by the five conditions.
The first three conditions are given by the saturation properties, 
saturation at normal nuclear density $\rho=\rho_0=0.168$ fm$^{-3}$,
the nuclear matter binding energy $B=-16$ MeV at saturation, 
and the nuclear incompressibility 
$K=9\rho^2\frac{\partial^2}{\partial\rho^2}\left(\frac{e}{\rho}\right)=380 \mathrm{MeV}$ (hard)
or $210~\mathrm{MeV}$ (soft).
Other two conditions come from the energy dependence of the optical potential.
The optical potential is required to take the values
\begin{align}
U_\mathrm{opt}(\rho_0,p=1.7\,\mathrm{GeV}) &=
U_\infty,\label{eq:optfix1}\\
U_\mathrm{opt}(\rho_0,p=p_0) &=0\, \mathrm{MeV},
\label{eq:optfix2}
\end{align}
where we adopt $U_\infty=60~\mathrm{MeV}$ and $p_0=0.65~\mathrm{GeV}$
in the nonrelativistic implementation.
Saturation condition
$P=\rho^2\frac{\partial}{\partial\rho}\left(\frac{e}{\rho}\right)=0$
and the 
saturation energy $e/\rho - m_N = B$
at 
$\rho=\rho_0$ leads to the Weisskopf relation 
\begin{equation}
\sqrt{m_N^2 + p_f^2} + \alpha + \beta + U_m(p_f) = m_N + B\,.
\label{eq:WR}
\end{equation}
We first fix the parameters in the momentum-dependent potential
and $U_0=\alpha+\beta$ by using Eq.~\eqref{eq:optfix1}--\eqref{eq:WR}.
Next the parameters of the density-dependent part ($\alpha, \beta$, and $\gamma$)
are fixed by using the saturation density and the incompressibility.
The details of the fitting procedure are found in Appendix~\ref{appendix:eos}.

We adopt a two-range Lorentzian-type momentum-dependent potential%
~\cite{Maruyama:1997rp,Isse:2005nk}
for the parameter sets MH2 and MS2
with the range parameters 
$\mu_1 = 2.02$ fm$^{-1}$ and $\mu_2=1.0$ fm$^{-1}$.
These parameters lead to $\gamma<2$, which has softer baryon density dependence than the hard EoS without a momentum-dependent part ($\gamma=2$) at high densities.
As an alternative parametrization, we assume one range Lorentzian  for
momentum-dependent potential, which leads to slightly harder EoS (MH1)
and softer EoS (MS1) at high densities.
Parameters 
are summarized in Table~\ref{table:ns}.
The effective mass for MH2 at the Fermi surface
is $m^*=p_F/v_F=p_F/\frac{\partial e}{\partial p}|_{p_F}=0.877m_N$
(formula given by Refs.~\cite{Danielewicz:1999zn,Jaminon:1989wj}),
while $m^*=0.705m_N$ for MH1.

\begin{figure*}[tbh]
\includegraphics[width=6.0cm]{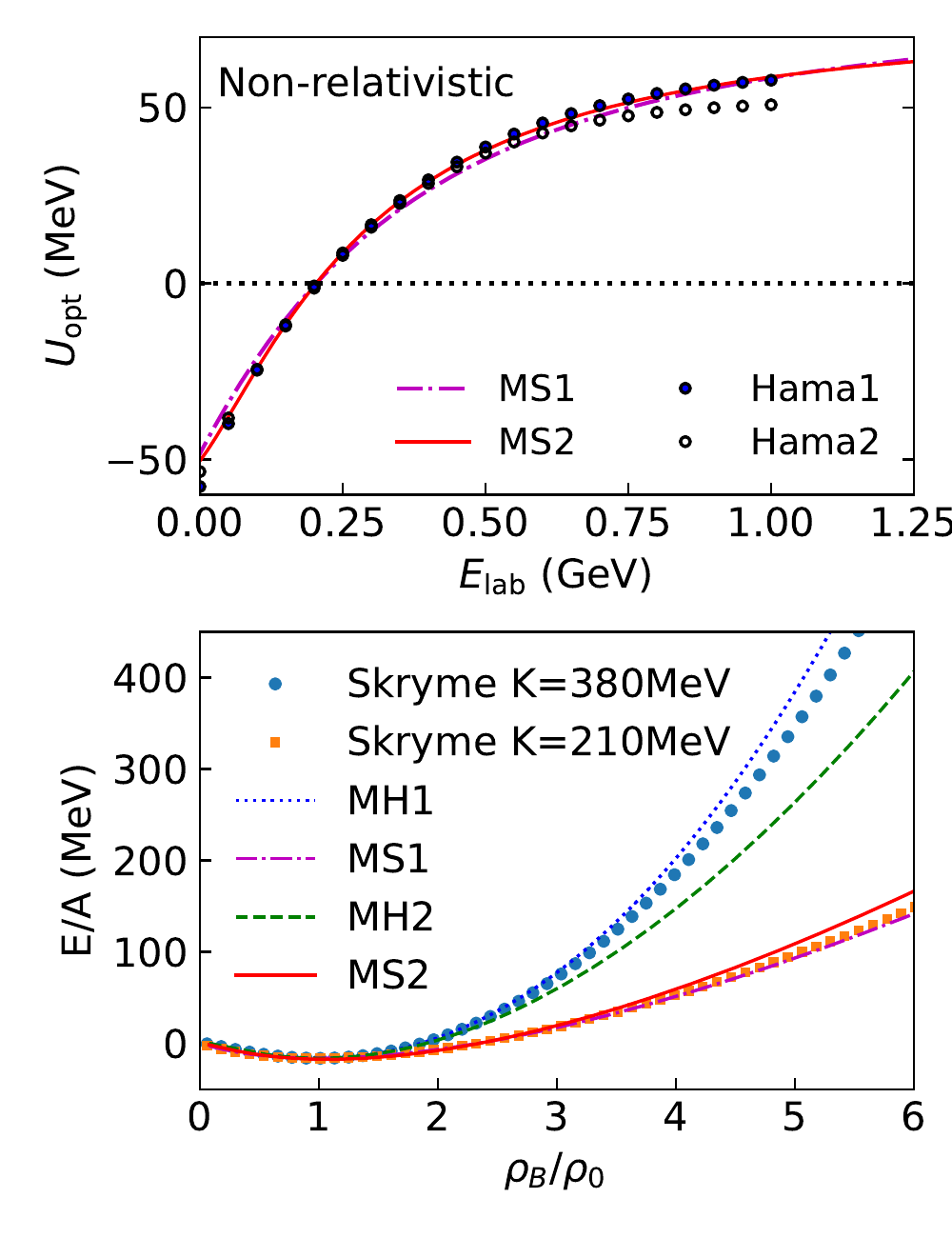}%
\includegraphics[width=6.0cm]{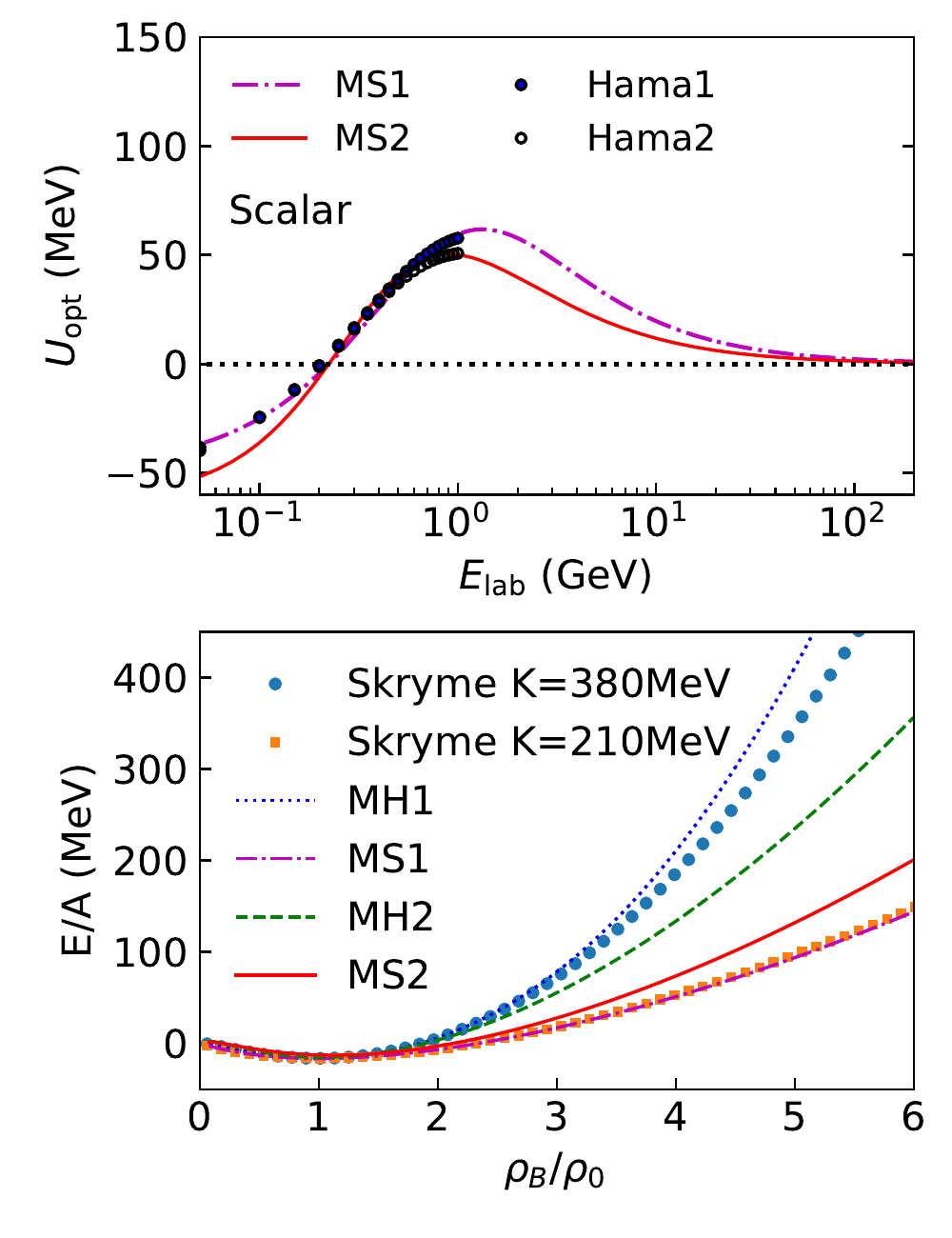}%
\includegraphics[width=6.0cm]{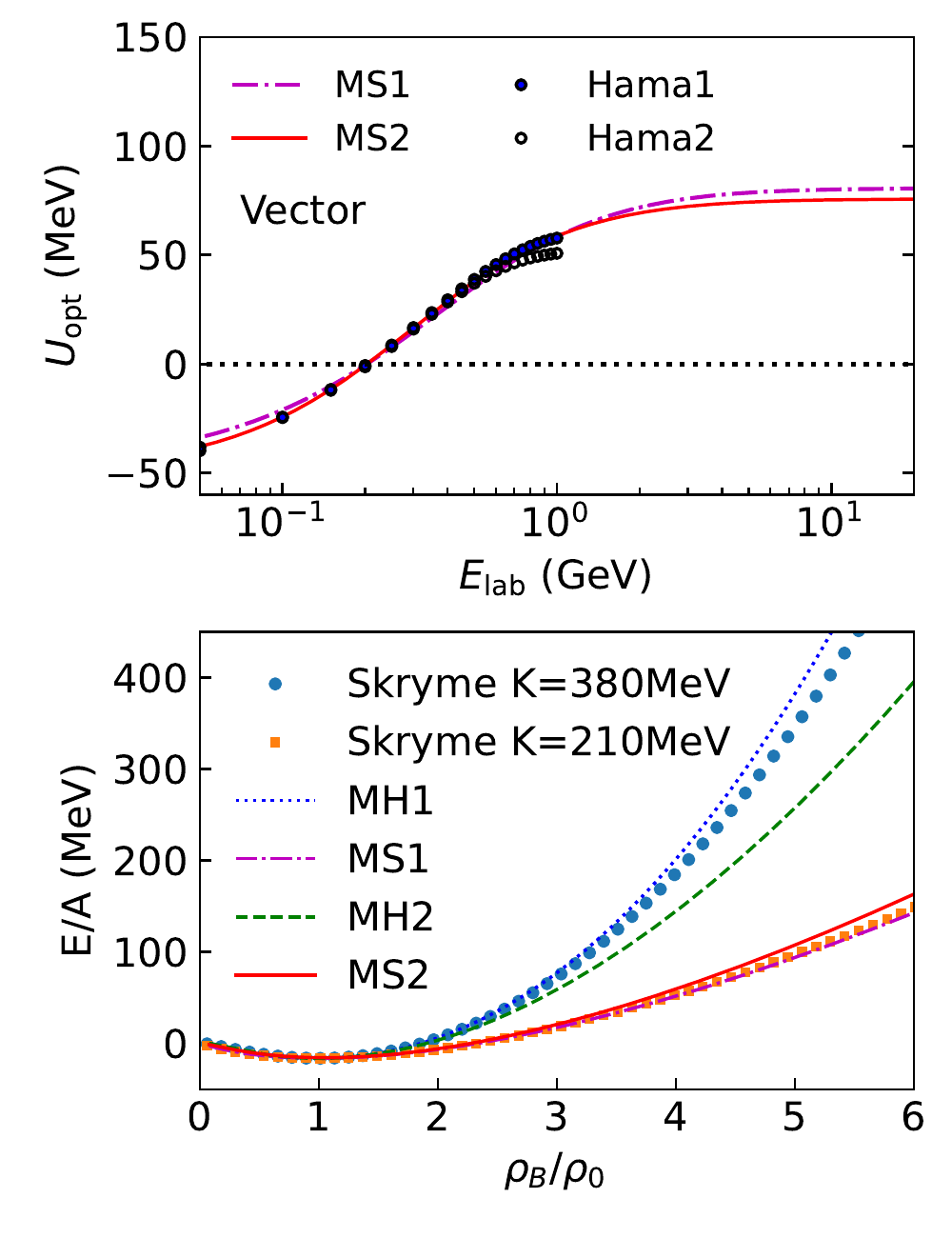}
\caption{
Incident energy dependence of the optical potential
in comparison with the real part of the global
Dirac optical potential~\cite{Hama:1990vr} (upper panels)
and the energy per nucleon as a function of the baryon density 
(normalized by $\rho_0$) (lower panels)
for different parametrizations.
Left, middle and right panels show the results
in the nonrelativistic potential treatment,
the scalar potential implementation,
and the vector potential implementation, respectively.
}
\label{fig:eos}
\end{figure*}
The left-upper panel of Fig.~\ref{fig:eos} compares 
the energy dependence of
the optical potential, Eq.(\ref{eq:opt}), with the real part of the global
Dirac optical potential~\cite{Hama:1990vr}.
Here we assume that the incident kinetic energy
$E_\mathrm{lab}=E-m_N$ is related
to the momentum in the argument of the potential by
the relation $(E-U_\mathrm{opt})^2-\bm{p}^2 = m_N^2$.
It is seen that all of the parameter sets (see Table~\ref{table:ns})
well describe the data except for the momentum-independent parametrization (H and S).

In the left-lower panel of Fig.~\ref{fig:eos}, 
the baryon-density dependence of the 
energy per nucleon
is shown for hard ($K=380$ MeV) and soft ($K=210$ MeV) EoSs.
Parameter set MH2 yields a softer EoS than the sets MH1.
This behavior can be also confirmed by comparing the values
of the $\gamma$ in Table~\ref{table:ns} as well as the
effective mass $m^*/m_N$.
When the effective mass at saturation density is smaller,
the EoS becomes harder at high densities.
The same trend is observed also in 
the relativistic mean field theory,
which demonstrates that the energy dependence of the optical potential 
modifies the EoS at high densities even if the incompressibility is fixed.

\subsection{Lorentz scalar potential}
\label{sec:rqmds}

In this section, we consider scalar potentials to construct EoS.
We assume the single-particle energy to be
\begin{equation}
 e^* = \sqrt{\bm{p}^2 + {m^*}^2},~~m^* = m + U_s + U_m
\end{equation}
The energy density of the nuclear matter
for the scalar potential is given by~\cite{RBUUp,Danielewicz:1999zn}
\begin{align}
 e &= \int d^3p \left( e^* - \frac{1}{2}\frac{m^*}{e^*}U_m(p)\right) f(x,p)
 \nonumber\\
  &- \rho_s U_s(\rho_s) + \int^{\rho_s}_0 U_s(\rho_s')d\rho_s',
\end{align}
where the scalar density is defined as
\begin{equation}
\rho_s = \int d^3 p \frac{m^*}{e^*}f(x,p).
\label{eq:sdensity}
\end{equation}
The density dependent scalar Skyrme potential is a function
of the scalar density, not the baryon density.
The momentum-dependent part of the potential takes the form
\begin{equation}
 U_m (\bm{p}) = \frac{C}{\rho_0}
     \int d\bm{p}'
    \frac{m^{*'}}{e^{*'}}
     \frac{f(x,p')}{1+[(\bm{p}-\bm{p}')/\mu]^2}.
    \label{eq:momdepScalar}
\end{equation}
The main difference between the previous and present approaches
is the form of prefactor $m^*/e^*$, which is also introduced
in the momentum-dependent potential.
A computation of the nuclear incompressibility $K$ for the scalar
 potential is given in Appendix~\ref{appendix:eos_scalar}.

In Table~\ref{table:ns}, we present EoS parameter sets
for scalar implementation of the Skyrme potential.
We compare the Dirac optical potential~\cite{Hama:1990vr}
with the single-particle energy of a nucleon,
subtracting kinetic energy
~\cite{Feldmeier:1991ey,Danielewicz:1999zn,Kapusta:2006pm}:
\begin{equation}
U_\mathrm{opt}(p) = e^* - \sqrt{m_N^2+\bm{p}^2}
\end{equation}
This optical potential approaches zero in the high momentum limit,
since the vector potential is not included.
As shown in the middle-upper panel of Fig.~\ref{fig:eos}, 
the optical potential vanishes at about $E_\mathrm{lab}\approx 100$ GeV.

We obtain the MH2 EoS which is slightly softer
than that in the nonrelativistic approach
when $\mu_1$ and $\mu_2$ are fixed to be 2.02 and 1.0 fm$^{-1}$.
We also provide MH1 and MS1 which have a closer
density dependence to the original hard and soft EoSs without
momentum dependence as depicted in the middle-lower panel of Fig.~\ref{fig:eos}.

\subsection{Lorentz vector potential}
\label{sec:rqmdv}

Next, we consider the vector implementation of
the Skyrme type density dependent mean field
and a Lorentzian type momentum-dependent mean field.  
The energy density of the nuclear matter
 has the following form~\cite{RBUUp}:
\begin{equation}
 e = \int d^3p \left( e^* + U_m^0 -
 \frac{1}{2}\frac{p^*_{\mu}}{e^*}U^\mu_m(p)\right) f(p)
  + \int_0^{\rho} U^0_\text{sk}(\rho')d\rho'
\end{equation}
where 
The momentum-dependent part of the potential takes the form:
\begin{equation}
 U_m^\mu (p)= 
     \frac{C}{\rho_0}
     \int d^3{p}'
     \frac{p^{*'\mu}}{e^{*'}}
     \frac{f(x,p')}{1+[(\bm{p}-\bm{p}')/\mu_k]^2},
     \label{eq:mdvec}
\end{equation}
We replace the argument of the momentum-dependent potential
with the relative momentum in the two-body center-of-mass (c.m.) frame
or the rest frame of a particle
in the  simulation.

We list our parameter set in Table~\ref{table:ns}
for the vector potential, too.
Energy dependence of the optical potential 
and baryon density dependence of the energy per nucleon 
are plotted in the right-bottom panel of Fig.~\ref{fig:eos}.

\section{Relativistic quantum molecular dynamics}
\label{sec:rqmd}
The RQMD model is a nonequilibrium
transport model which can simulate the space-time evolution of
the $N$ particles interacting via the potentials
based on the constrained Hamiltonian dynamics~\cite{Komar:1978hc}.

In this section, 
we first give a brief explanation of basics in the RQMD model.
Next we give the equations of motion for the RQMD/S
approach with the nonrelativistic potential implementation,
and then present the equation of motion for RQMD with
scalar-vector implementation of potentials
in the on-mass-shell constraints.

\subsection{Preliminaries}

In the RQMD approach, $8N$ phase space variables
are reduced by the $2N$ constraints to obtain the physical $6N$ phase space, 
$\phi_i\approx 0\,(i=1,\ldots, 2N)$
with $\approx$ representing the weak equality satisfied on the realized evolution path.
According to Dirac's constraint Hamiltonian formalism,
the Hamiltonian is given by a linear combination of $2N-1$ constrains,
\begin{equation}
  H = \sum^{2N-1}_{i=1} u_i \phi_i.
\label{eq:RQMD}
\end{equation}
Among the $2N$ constraints, 
the first $N$ ($i=1\ldots N$) are the on-mass-shell constraints
and the latter $N$ ($i=N+1\ldots 2N$) represent the time fixation of the particles. Since one of the time fixation constraints defines
the evolution temporal parameter $t$, we use $2N-1$ constraints in Eq.~\eqref{eq:RQMD}.
The equations of motion are 
\begin{align}
  \frac{dq_i^\mu}{dt} =[H, q^{\mu}_i]
  \approx \sum_{j=1}^{2N-1} u_j \frac{\partial \phi_j}{\partial p_{i\mu}},\\
  \frac{dp_i^\mu}{dt} =[H, p_i^\mu]
  \approx -\sum_{j=1}^{2N-1} u_j \frac{\partial \phi_j}{\partial q_{i\mu}},
\end{align}
where the Poisson brackets are defined as
\begin{equation}
  [A, B] = \sum_{k}\left(
   \frac{\partial A}{\partial p_k^\mu}
   \frac{\partial B}{\partial q_{k\mu}}
 - \frac{\partial A}{\partial q_k^\mu}
   \frac{\partial B}{\partial p_{k\mu}}
      \right).
\end{equation}

The Lagrange multipliers ($u_i$) are determined by requiring that the constraints remain satisfied.
For $i=1\ldots 2N-1$, $\phi_i$ is assumed not to explicitly contain
the evolution time parameter $t$;
then we find
\begin{align}
\frac{d\phi_i}{dt} =&[H,\phi_i]
=\sum_{j=1}^{2N-1} C_{i,j}u_j 
\approx 0 ,\\
C_{i,j}=& [\phi_j,\phi_i].
\end{align}
By comparison, $\phi_{2N}$ is assumed to explicitly contain $t$,
\begin{align}
\frac{d\phi_{2N}}{dt}
=&[H,\phi_{2N}]+\frac{\partial \phi_{2N}}{\partial t}
\nonumber\\
=&\sum_{j=1}^{2N-1} C_{2N,j}\,u_j
+\frac{\partial \phi_{2N}}{\partial t} \approx 0.
\end{align}
Thus, by defining $u_{2N}=0$, the Lagrange multipliers are obtained by solving the following linear equation:
\begin{align}
&    \sum_{j=1}^{2N} C_{i,j} u_j 
    = -\delta_{i,2N} \frac{\partial \phi_{2N}}{\partial t},
\\
&\to u_i = - C^{-1}_{i,2N} \frac{\partial \phi_{2N}}{\partial t}.
\end{align}

We adopt the time fixation constraints
\begin{align}
\phi_{i+N}=&\hat{a}\cdot(q_i-q_N)\approx 0\ (i=1\ldots N-1),
\label{eq:time1}\\
\phi_{2N}=&\hat{a}\cdot q_N - t \approx 0.
\label{eq:time2}
\end{align}
where $\hat{a}= P/\sqrt{P^2}$, and
$P=\sum_{i=1}^np_i$ is the total momentum~\cite{Marty:2012vs}.
In this choice, the time coordinates of
all particles becomes the same in the overall center-of-mass system
as $\hat{a}$ becomes the unit vector $(1,\bm{0})$ in this frame,
which allows us to obtain the Lagrange multipliers analytically
~\cite{Maruyama:1996rn}.
By replacing $p_i^0$ in the potential with the kinetic energy,  $\sqrt{\bm{p}_i^2+m^2}$,
the matrix $C_{i,j}$ and its inverse $C^{-1}_{i,j}$ are found to have the form
\begin{align}
    C_{i,j}=\begin{pmatrix} * & -D \\ D^T & 0 \end{pmatrix},
    C^{-1}_{i,j}=\begin{pmatrix} * & (D^{-1})^T \\ -D^{-1} & 0 \end{pmatrix}.
\end{align}
In the overall center-of-mass system,
the matrix $D$ is obtained as $D_{i,j}=[\phi_i,\phi_{N+j}]=\partial\phi_i/\partial p_i^0-\delta_{iN}\partial\phi_N/\partial p_N^0$ ($j=1\ldots N-1$)
and $D_{i,N}=\delta_{iN}\partial\phi_N/\partial p_N^0$.
In the case where the on-mass-shell constraint is given as
$\phi_i=p_i^2-m_i^2-F_i(p,q)$ 
with $F_i$ representing the potential effects,
the inverse of $D$ matrix is found to be
$D^{-1}_{i,j}=\delta_{ij}(1-\delta_{iN})/2p_i^0+\delta_{i,N}/2p_N^0$.
Then the Lagrange multiplier is found to be
\begin{align}
    u_i =& -\frac{1}{2p_i^0}\,\frac{\partial\phi_{2N}}{\partial t}
= \frac{1}{2p_i^0}
\ (i=1\ldots N),
\\
u_i=&0\ (i=N+1\ldots 2N-1).
\end{align}

In the following subsections,  
we compare the results of on-mass-shell constraints in the nonrelativistic, scalar, and vector implementations of the potentials.

\subsection{Equations of motion for the RQMD/S model}
\label{sec:simplifiedrqmd}

In this section, we present the equations of motion
for the RQMD/S model~\cite{Isse:2005nk}.
In the RQMD/S model,
the EoS in the nonrelativistic implementation is used,
and the on-mass-shell constraint is given as
$\phi_i=p_i^2-m_i^2-2m_iV_i\approx 0$.
Thus the one-particle energy for the $i$th particle takes the form
\begin{equation}
 p^0_i = \sqrt{\bm{p}_i^2 + m_i^2 + 2m_i V_i}.
 \label{eq:scalar1}
\end{equation}

Then the above ansatz leads to the equations of motion for $N$-particle system in the RQMD/S approach~\cite{Maruyama:1996rn}
\begin{equation}
\frac{dq^\mu_i}{dt} = \frac{p^\mu_i}{p^0_i} - \sum_{j=1}^N
  \frac{m_j}{p^0_j}\frac{\partial V_j}{\partial p_{i\mu}},~~~
\frac{dp^\mu_i}{dt} = \  \sum_{j=1}^N
  \frac{m_j}{p^0_j}\frac{\partial V_j}{\partial q_{i\mu}}.
\label{eq:motion1}
\end{equation}
The suppression factor $m_i/p^0_i$ appearing in the equations of motion 
is the direct consequence of scalar implementation
of the potential in Eq.~(\ref{eq:scalar1}).
Furthermore, we make an ansatz that 
the potential of the $i$th particle in RQMD/S depends on
the scalar density of the form~\cite{RQMD1989}
\begin{equation}
 \rho_i = \sum_{j\ne i}^N \rho_{ij} 
 = \sum_{j\ne i}^N 
 \frac{1}{(4\pi L)^{3/2}}
    \exp\left( \frac{q_{Tij}^2}{4L} \right)\,,
    \label{eq:scalarDensityRQMDS}
\end{equation}
where $q_{Tij}=q_{ij}-(q_{ij}\cdot u_{ij}) u_{ij}$ is the distance
in the two-body center-of-mass frame of particles $i$ and $j$,
where $q_{ij}=q_i-q_j$ and $u_{ij}=(p_i+p_j)/\sqrt{(p_i+p_j)^2}$.

The one-particle potential $V_i$ in the RQMD/S framework is not really the single particle potential, but the sum of $V_i$ gives the total potential energy in the nonrelativistic limit.
Thus we need to consider the $i$th particle contribution to the total potential energy
given in Eq.~\eqref{eq:ematter}.
By assuming the Gaussian profile in the QMD approach
\begin{equation}
\tilde{\rho}_i(\bm{r})=\frac{1}{(2\pi L)^{3/2}}
  \exp\left(-\frac{(\bm{r}-\bm{r}_i)^2}{2L}\right)\,,
\end{equation}
the scalar density Eq.~\eqref{eq:scalarDensityRQMDS} is obtained as the sum of
density overlap $\rho_{ij}$,
\begin{align}
\langle \rho_i \rangle &\equiv \sum_{j(\neq i)}\int d^3r \tilde{\rho}_i(\bm{r})\tilde{\rho}_j(\bm{r})
 = \sum_{j(\neq i)} \rho_{ij} = \rho_i \,,\\
\rho_{ij} &= \frac{1}{(4\pi L)^{3/2}}
\exp\left[-\frac{(\bm{r}_i-\bm{r}_j)^2}{4L}\right]\,.
\end{align}
Note that $q_{Tij}^2=-(\bm{r}_i-\bm{r}_j)^2$
in the center-of-mass frame of $i$th and $j$th particles.
We further assume that the momentum spread of the wave packet is small enough.
Then we obtain the total potential energy is given as the sum of $i$th particle contributions,
\begin{align}
 V = \sum_{i}V_i
 &\simeq \sum_i \Big( \frac{\alpha}{2\rho_0} \rho_i
         + \frac{\beta}{(\gamma+1)\rho_0^\gamma} \rho_i^\gamma \nonumber\\
    &+\frac{C}{2\rho_0}
    \sum_{j(\ne i)}\frac{\rho_{ij}}{1+(\bm{p}_i-\bm{p}_j)^2/\mu^2}
             \Big).
             \label{eq:RQMDpot}
\end{align}
We have made an approximation for the second term:
\begin{equation}
\int d^3r \left(\sum_i \tilde{\rho}_i(r) \right)^{\gamma+1}
\simeq \sum_i  \left( \sum_{j(\neq i)} \int d^3r \tilde{\rho}_i(r) \tilde{\rho}_j(r) \right)^\gamma
\end{equation}
This is an approximation adopted in most of the QMD codes.
It is possible to change the width for calculating this repulsive part to
justify this approximation~\cite{Maruyama:1997rp,Maru1990}.
Since our purpose in this paper is not an extraction of the equation of state,
we do not make such a modification.
Accordingly, the density and momentum dependence of
the one-particle potential for the $i$th particle in the RQMD/S approach
$V_i = V_{\mathrm{sk},i} + V_{m,i}$
are taken to be
\begin{align}
 V_{\mathrm{sk},i} &=
 \frac{\alpha}{2\rho_0} \rho_i + \frac{\beta}{(1+\gamma)\rho_0^\gamma}
  \rho_i^\gamma ,\\
 V_{m,i} &= \sum_{j\ne i}^N V_{m,ij}\rho_{ij},~~
 V_{m,ij} =\frac{C}{2\rho_0} \frac{1}{1-(p_{Tij}/\mu)^2}\,.
\end{align}
We also use the relative momentum $p_{T,ij}=p_{ij}-(p_{ij}\cdot u_{ij})u_{ij}$
in the center-of-mass frame of two particles
in the momentum-dependent potential, 
where $p_{ij}=p_i - p_j$.
In Appendix~\ref{sec:correction}, we present corrections to the previous paper~\cite{Isse:2005nk} regarding the implementation of these equations of motion in the code.

We now argue that the above potential energy plays the role of
the potential energy in Hamilton's equations of motion.
The nonrelativistic limit was checked in Ref.~\cite{RQMD1989,Isse:2005nk}.
Here, we follow a different way.
In order to take account of the relativistic effects,
it is useful to consider the following Hamiltonian~\cite{Maruyama:1996rn}:
\begin{equation}
 \mathcal{H}
   =\sum_{i=1}^N E_i\,,\ 
 E_i=\sqrt{\bm{p}_i^2 + m_i^2+ 2m_i V_i},
\end{equation}
The usual Hamilton equations of motion
from $\mathcal{H}$ are
the same as the spatial part of the RQMD/S equations of motion
Eq.~\eqref{eq:motion1} in the overall center-of-ass system
after substituting the on-mass-shell constraint, $p_i^0=E_i$.
Therefore the Hamiltonian dynamics using $\mathcal{H}$
is found to be equivalent to RQMD/S,
which is a framework of the Dirac's constrained Hamiltonian dynamics
with the on-mass-shell constraint and the global time-fixation constraint
in the center-of-mass system.
The nonrelativistic limit is obvious
\begin{equation}
 \mathcal{H} 
   \simeq \sum_{i=1}^N \left( \frac{\bm{p}_i^2}{2m_i} + V_i + m_i \right)
\,.
\end{equation}
From these discussions, the potential energy in the RQMD/S
treatment is equivalent to the mean-field potential energy
in the nonrelativistic limit.
We note that a similar approach has been discussed
in the framework of RBUU in Ref.~\cite{RBUUp}.

\subsection{The equations of motion for scalar-vector potentials}

We present both scalar and vector implementation of
above phenomenological potentials within the framework
of the RQMD approach~\cite{RQMD1989}.

We impose on-mass shell condition:
\begin{equation}
  H_i = p_i^{*2}-m_i^{*2}
      = (p_i - V_i)^2 - (m_i -S_i)^2=0
      \label{eq:onmass}
\end{equation}
for $i$th particles, where $V^\mu_i$ and $S_i$ are
the one-particle vector and scalar potentials,
together with the time fixation constraints give by
Eqs.~(\ref{eq:time1}) and (\ref{eq:time2}).
Then, the equations of motion are given by
\begin{align}
\frac{dq^\mu_i}{dt} & =
  2u_i p_i^{*\mu}
     -2\sum_{j=1}^N u_j\left(
      {m_j^*}
      \frac{\partial m_j^*}{\partial p_{i\mu}}
      +p^{*\nu}_j
      \frac{\partial {V}_{j\nu}}{\partial p_{i\mu}}
       \right),\\
\frac{dp^\mu_i}{dt} &=
      2\sum_{j=1}^N u_j\left(
      m_j^*
      \frac{\partial m_j^*}{\partial q_{i\mu}}
      +p^{*\nu}_j
     \frac{\partial V_{j\nu}}{\partial q_{i\mu}}
      \right).
     \label{eq:eomrqmd}
\end{align}
We need to invert numerically an $N\times N$ matrix to obtain the Lagrangian multipliers
$u_i$ at each time step.
Within those constraints together with the assumption that
the arguments of the potentials are replaced by the free one,
the Lagrangian multipliers become $u_i=1/2p_i^{*0}$
in the overall center-of-mass frame.
Then one obtains the equations of motion for $i$the particle as
\begin{align}
\frac{dq^\mu_i}{dt} & =
   \frac{p_i^{*\mu}}{p_i^{*0}}
     -\sum_{j}\left(
      \frac{m_j^*}{p_j^{*0}}
      \frac{\partial m_j^*}{\partial p_{i\mu}}
      +v^{*\nu}_j
      \frac{\partial {V}_{j\nu}}{\partial p_{i\mu}}
       \right), \nonumber\\
\frac{dp^\mu_i}{dt}
      &= \sum_{j}\left(
      \frac{m_j^*}{p_j^{*0}}
      \frac{\partial m_j^*}{\partial q_{i\mu}}
      +v^{*\nu}_j
     \frac{\partial V_{j\nu}}{\partial q_{i\mu}}
      \right),
\end{align}
where $v_i^{*\mu}=p_i^{*\mu}/p_i^{*0}$.
The equations of motion for the kinetic momentum 
$\bm{p}_i^*= \bm{p}_i - \bm{V}_i$ may be obtained by adding the
derivative of the vector potential:
\begin{equation}
 \dot{V}^\mu_i  = 
    \sum_j\left( 
    \dot{x}^\nu_j \frac{\partial V^\mu_i}{\partial x^\nu_j}
    +
    \dot{p}^\nu_j \frac{\partial V^\mu_i}{\partial p^\nu_j}
    \right).
\end{equation}
The explicit form of the equations of motion for both RQMDs and RQMDv
can be found in Appendix~\ref{appendix:eom_rqmd}.


In RQMD, the one-particle potentials $S_i$ and $V^\mu_i$
 are dependent on the scalar density
$\rho_{si}$ and the baryon current $J^\mu_i$, respectively,
which are obtained by
\begin{equation}
\rho_{s,i}=\sum_{j\neq i} \frac{m^*_j}{p_j^{*0}}\rho_{ij},~~
J^\mu_i=\sum_{j\neq i}B_j v^{*\mu}_j \rho_{ij}
\label{eq:RQMDdensity}
\end{equation}
where $B_j$ is the baryon number of the $j$th particle,
and $\rho_{ij}$ is the so-called interaction density
(overlap of density with another hadron wave packet)
which will be specified below.
The vector potential is defined by using the baryon current 
~\cite{Danielewicz:1998pb,Sorensen:2020ygf}
\begin{equation}
V^\mu_i = B_i\frac{V_{\mathrm{sk},i}(\rho_{Bi})}{\rho_{Bi}} J_i^\mu,
\label{eq:vectorpotential}
\end{equation}
where $\rho_{Bi}=\sqrt{J_i^\mu J_{i\mu}}$ is the invariant baryon density.
The momentum-dependent part of the one-particle potential in the vector implementation is given by
\begin{equation}
 V_{mi}^\mu (p_{Tij})=\sum_{k=1,2}\frac{C_k}{2\rho_0}
 \sum_{j\neq i}
     \frac{p_j^{*\mu}}{p_j^{*0}}
     \frac{\rho_{ij}} {1-[p_{Tij}/\mu_k]^2}
     \label{eq:MVpotV}
\end{equation}
while the scalar implementation is
\begin{equation}
 V_{mi} (p_{Tij})=\sum_{k=1,2}\frac{C_k}{2\rho_0}
 \sum_{j\neq i}
     \frac{m_j^{*}}{p_j^{*0}}
     \frac{\rho_{ij}} {1-[p_{Tij}/\mu_k]^2},
     \label{eq:MVpotS}
\end{equation}
where $p_{Tij}$ is the relative momentum between particle $i$ and $j$,
which will be specified below.
As the equations of motion are obtained by assuming that the argument of
the potentials is replaced by the free one, we also replace 
$m^*_j$ and $p^{*\mu}$ in the definition of the scalar density and baryon
current as well as the momentum-dependent potentials with the free one.

We now discuss the form of the interaction density in the RQMD approach.
As a first option, we use the following interaction density:
\begin{equation}
 \rho_{ij}=\frac{\gamma_{ij}}{(4\pi L)^{3/2}}\exp(q^2_{Tij}/4L)\,,
 \label{eq:twobody}
\end{equation}
where $q_{Tij}$ is the distance in the center-of-mass frame of
the particles $i$ and $j$, 
\begin{eqnarray}
q_{Tij} &=& q_{ij} - (q_{ij}\cdot u_{ij})u_{ij},~~
 u_{ij}=P_{ij}/\sqrt{P_{ij}^2} \label{eq:twobodydistance},\\
q_{ij} &=& q_i - q_j,~~ P_{ij}=p_i+p_j
\end{eqnarray}
and $\gamma_{ij}=P_{ij}^0/\sqrt{P_{ij}^2}$ is the Lorentz $\gamma$ factor
to ensure the correct normalization of the Gaussian%
~\cite{Oliinychenko:2015lva}.
We obtain the RQMD/S model which follows the original RQMD~\cite{RQMD1989,RQMDmaru},
by replacing the normalization factor $\gamma_{ij}$ with $\gamma_j=p^0_j/m_j$
to obtain Eq.~(\ref{eq:scalarDensityRQMDS}), which is the Lorentz scalar.
With this replacement, we lose a correct normalization of the Gaussian,
but this would not be a problem as one may  adjust the width parameter of the Gaussian because we have only scalar potentials.
The same approximation was used in Ref.~\cite{Fuchs:1996uv}
for both the scalar density and the baryon current for low energy heavy-ion collisions, $E_\mathrm{lab}< 2A$ GeV.
We found that this approximation overestimates
the vector density significantly at relativistic energies. 
Thus, the predictions from this approach are
not reliable at relativistic energies.

Another approach is to use the rest frame of a particle $j$,
\begin{equation}
 q^2_{R,ij} = (q_i - q_j)^2 - [(q_i - q_j)\cdot u _j]^2,~~
 u_i = p_j/m_j
 \label{eq:restframedistance}
\end{equation}
for the definition of the two-body distance in the argument of the potential.
This is used in the relativistic Landau-Vlasov model~\cite{RLV}, in which
Gaussian shape is used to solve the relativistic Boltzmann-Vlasov equation.
In this case, the interaction density takes the form
\begin{equation}
 \rho_{ij}=\frac{\gamma_{j}}{(4\pi L)^{3/2}}\exp(q^2_{R,ij}/4L)\,,
 \label{eq:restframe}
\end{equation}
where $\gamma_j=p^0_j/m_j$,
When we substitute this interaction density into Eq.~(\ref{eq:RQMDdensity}),
the Lorentz factor in front of the Gaussian cancels the factor in the scalar density,
and the scalar density becomes manifestly Lorentz scalar, and the baryon current is also covariant vector without loss of the correct normalization of the Gaussian:
\begin{equation}
 \rho_{s,i}=\sum_{i\neq j}\hat{\rho}_{ij},~~ 
 J^\mu_{i}=\sum_{i\neq j} B_j u_j
 \hat{\rho}_{ij}
\end{equation}
where $\hat{\rho}_{ij}=(1/4\pi L)^{3/2}\exp[q_{R,ij}^2/4L]$,
and $u_j$ is a four-velocity: $u_j=\gamma_j(1,\bm{p}_j/p^{*0}_j)=(\gamma_j, \bm{p}_j/m_j)$.
The derivatives of the transverse distances can be found
in Appendix~\ref{appendix:derivative}.

Numerically, the main difference between the two-body distance at the c.m. of two particles
$q_{T,ij}$ and the rest frame of a particle $q_{R,ij}$
in the estimation of the interaction density
$\rho_{ij}$ is using the different shapes of the Gaussian.
So we expect that if a violation of the Lorentz invariance is not significant,
two different choices  may yield the same results with a possible different choice of the Gaussian width. Numerical study indicates that Eq.~(\ref{eq:restframe}) needs generally 
smaller Gaussian width than Eq.~(\ref{eq:twobody}) to obtain
a similar result at ultrarelativistic energies, as shown below.

\subsection{Numerical implementation}

The mean-field models mentioned above have been implemented in the JAM2
Monte Carlo event generator.
The physics of the collision term in JAM2 is the same as in the previous version of JAM~\cite{JAMorg}, in which particle productions are modeled by the resonance 
(up to 2 GeV) and string excitations and their
decays~\cite{Sorge:1995dp,UrQMD1,UrQMD2,GiBUU,SMASH}.
The leading hadron that contains original constituent quarks
can scatter again with reduced cross section within its formation time
to simulate effectively the quark interactions.
There are several improvements: We use Pythia8 event generator~\cite{Pythia8}
to perform string decay as well as the hard scatterings instead of Pythia6~\cite{Pythia6}.
Resonance excitation cross sections are also changed to improve the threshold
behavior by fitting the matrix elements~\cite{UrQMD1,UrQMD2,GiBUU,SMASH}.
As technical improvements in JAM2,
we introduced expanding boxes to reduce the computational time
for both two-body collision term and potential interaction.
A detailed explanation will be presented elsewhere.

\section{Results for anisotropic flows}
\label{sec:results}

We consider three types of anisotropic flows. The first one is the sideward flow
$\langle p_x \rangle$, which is the mean particle transverse momentum
projected on to the reaction plane, where angle brackets indicate an average over particles and events.
The directed flow $v_1$ is also used, which is defined by 
\begin{equation}
v_1=\langle \cos\phi \rangle = \left\langle \frac{p_x}{p_T} \right\rangle,
\end{equation}
where $\phi$ is measured from the reaction plane, and $p_T=\sqrt{p_x^2+p_y^2}$
is the transverse momentum. The $z$ axis is the beam direction.
The elliptic flow 
\begin{equation}
v_2=\langle \cos 2\phi\rangle
= \left\langle \frac{p_x^2 - p_y^2}{p_T^2} \right\rangle
\end{equation}
reflects the anisotropy of transverse particle emission.
These anisotropic flows are sensitive to the pressure built up during the
collisions and thus sensitive to the mean field in the microscopic transport models~\cite{Danielewicz:1998vz,Danielewicz:1998pb}.

It was shown in Ref.~\cite{Hartnack:1997ez}
that the sideward flow is sensitive to the Gaussian width parameter
for heavy-ion collisions in the $E_\mathrm{lab}\approx 1A$ GeV regime,
because the Gaussian width $L$ controls the range and strength
of the interaction in the QMD approach.
Thus, we will examine width dependence of the flow.
IQMD uses $L=2.165$ fm$^2$ for the Au nucleus to obtain a stable nuclear density profile~\cite{Hartnack:1997ez}, while UrQMD~\cite{UrQMD1} uses $L=1.0$ fm$^2$.
The width in the JQMD model~\cite{Niita:1995mc} is $L=2.0$ fm$^2$.
A recent QMD model called PHQMD~\cite{Aichelin:2019tnk} uses $L=0.54$ fm$^2$.

We first compare RQMD/S with the relativistic quantum molecular dynamics
with the scalar potential (RQMDs).
Then, new results from the relativistic quantum molecular dynamics with
the vector potential (RQMDv) will be present.

\subsection{Comparison of RQMD/S and RQMDs}

In this section, we compare the results from two different RQMD models with scalar potentials.
The name RQMDs1 is used when the two-body distance Eq.~(\ref{eq:twobodydistance}) is used for the argument of potential,
similarly RQMDs2 for Eq.~(\ref{eq:restframedistance}).

\begin{figure*}[hbt]
\includegraphics[width=8.0cm]{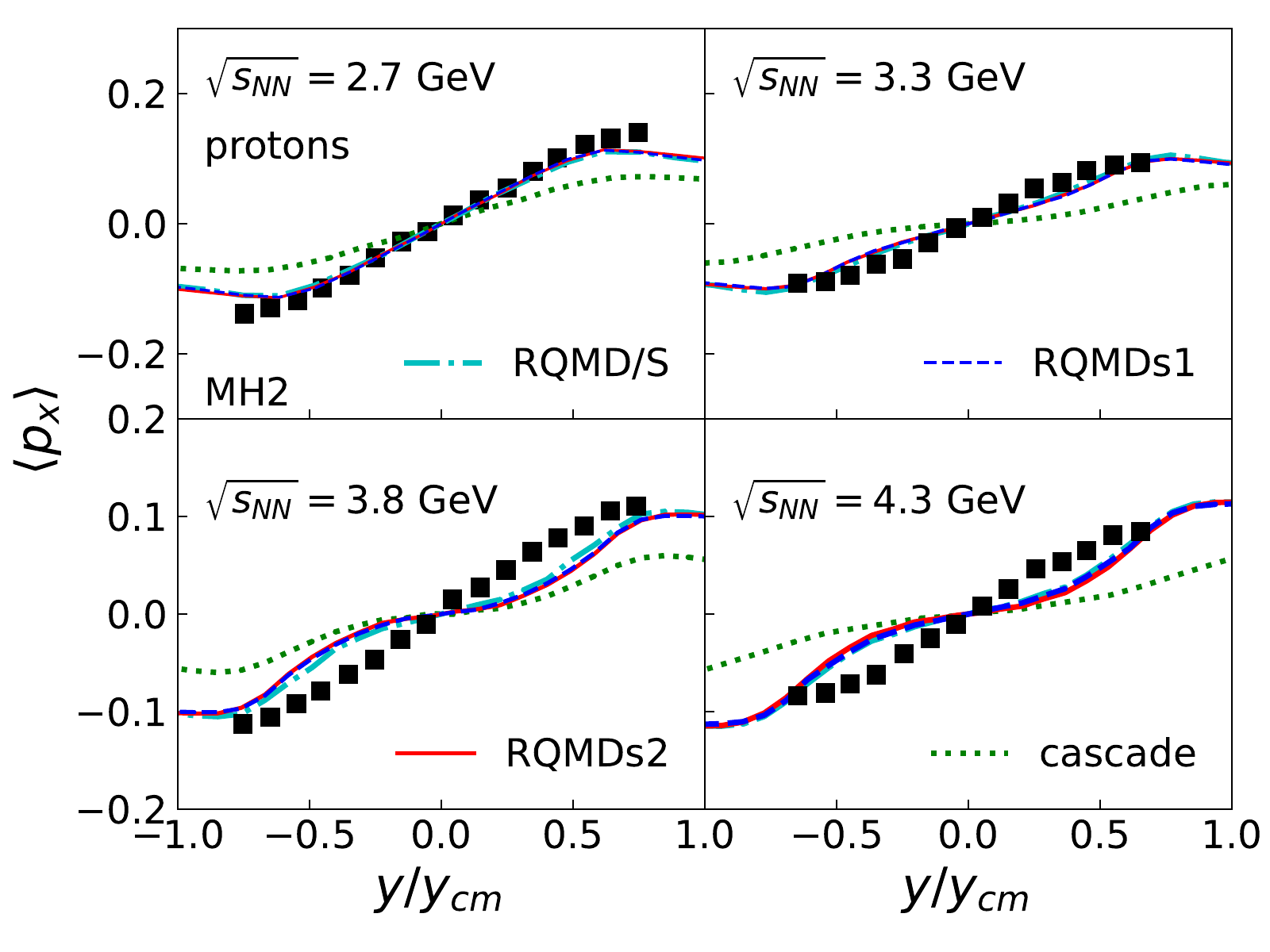}%
\includegraphics[width=8.0cm]{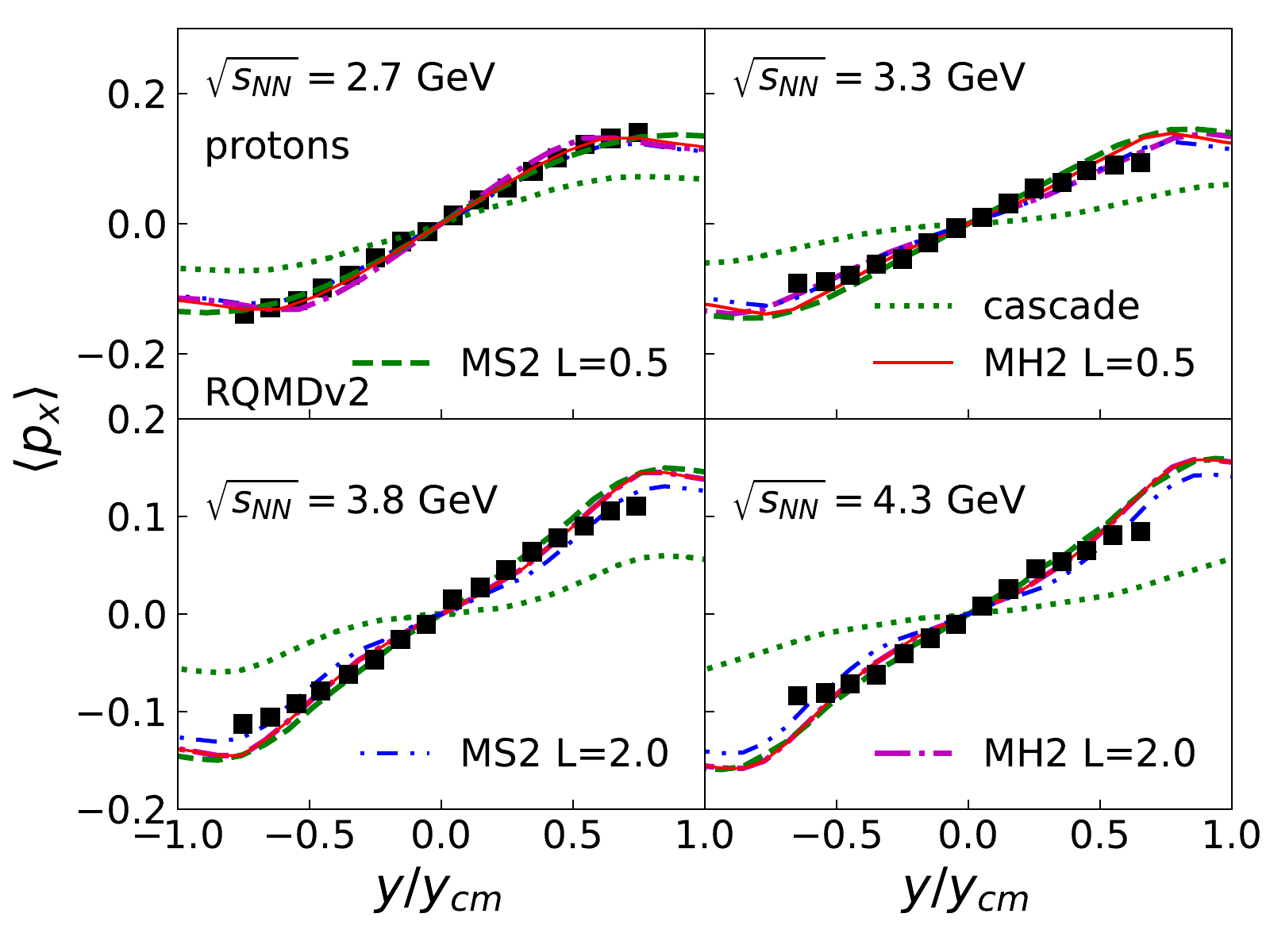}
\caption{The rapidity dependence of the sideward flow.
in comparison with mid-central Au + Au collisions at $\sqrt{s_{NN}}=2.7, 3.3, 3.8$, and 4.3 GeV
from the E895 data~\cite{E895v1}.
In the left panel, we show the results 
from the cascade mode (dotted lines), 
RQMD/S (dotted-dashed lines), RQMDs1 (dashed lines), and RQMDs2 (solid lines).
Momentum-dependent hard EoS (MH2) is used in the calculations.
In the right panel, we show the sideward flow from RQMDv2.
RQMDv2 results for the EoS MS2 with $L=0.5$ fm$^2$
are represented by dashed lines,
MH2 with $L=0.5$ fm$^2$ by solid lines,
MS2 with $L=2.0$ fm$^2$ by double-dotted lines,
and MH2 with $L=2.0$ fm$^2$ by dashed-dotted lines.
}
\label{fig:pxRQMD}
\end{figure*}

In the left panel of Fig.~\ref{fig:pxRQMD},
we compare the proton sideward flow $\langle p_x\rangle$
in mid-central Au + Au collisions at 
$\sqrt{s_{NN}}=2.7, 3.3, 3.8$, and 4.3 GeV
($E_\mathrm{lab}=1.85, 4, 6, 8A$ GeV).
We use $L=2.0$ fm$^2$ for the Gaussian width
as used in the previous calculations~\cite{Isse:2005nk}.
The impact parameter range is chosen to be $4< b< 8$ fm.
In the left panel of Fig.~\ref{fig:pxRQMD}, we compare four different approaches:
RQMD/S, RQMDs1, and RQMDs2 for the MH2 EoS, and cascade mode, in which
only the collision term is included and potentials are disabled.
As is well known, the cascade model lacks some pressure at AGS energies
($2.3<\sqrt{s_{NN}}<5$ GeV)
and significantly underestimates the sideward flow.
Both RQMD/S and RQMDs improve the description of the sideward flow due to an additional pressure generated by the mean-field.
However, all calculations with scalar potentials predict less flow compared to the experimental data.
 
All three models with scalar potentials show good agreement with each other.
The agreement of RQMDs with RQMD/S may justify the approximations in the RQMD/S model, which significantly simplifies the model compared to RQMDs.

We do not show the results for other parameter sets MS2, MH1, MS1
because their results are almost identical to those from MH2 in RQMD/S and RQMDs.
This insensitivity of the sideward flow to the EoS
is consistent with our previous finding in Ref.~\cite{Isse:2005nk}.
We argue that the main reason for this insensitivity is to use the scalar potential,
which is a function of the scalar density.
We have checked that the sideward flow results
are not significantly modified with a smaller width $L=0.5$ fm$^2$ at AGS energies in the RQMDs model.

In the previous paper~\cite{Isse:2005nk}, we showed that RQMD/S reproduces the flow data with the momentum-dependent mean-field.
The reason for the discrepancy between the current result and the previous one is that we overestimate force by a factor of 2 due to a mistake in the earlier calculations. A comparison of the previous one and the corrected one is provided in Appendix~\ref{sec:correction}.

\begin{figure*}[tbh]
\includegraphics[width=7.5cm]{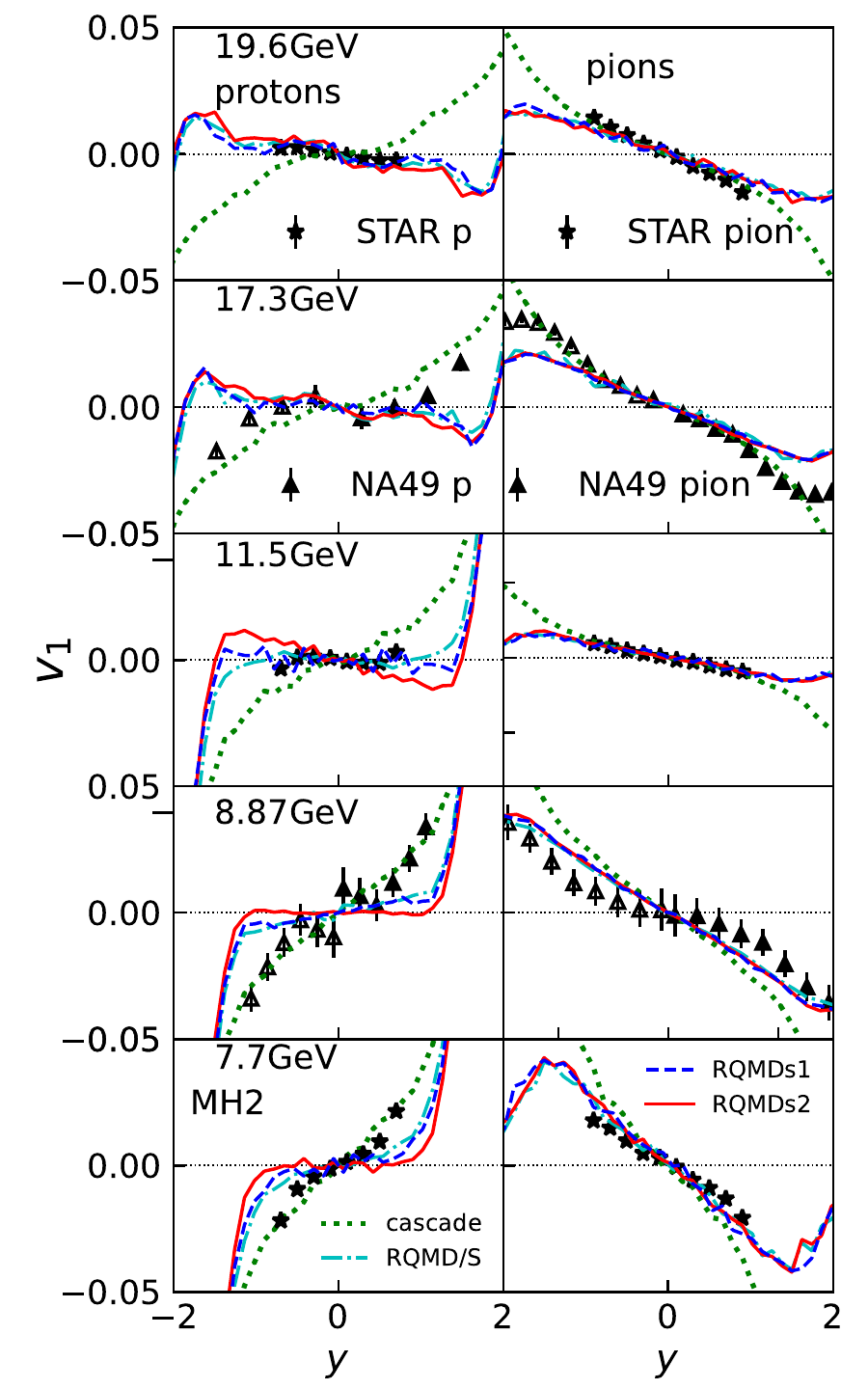}%
\includegraphics[width=7.5cm]{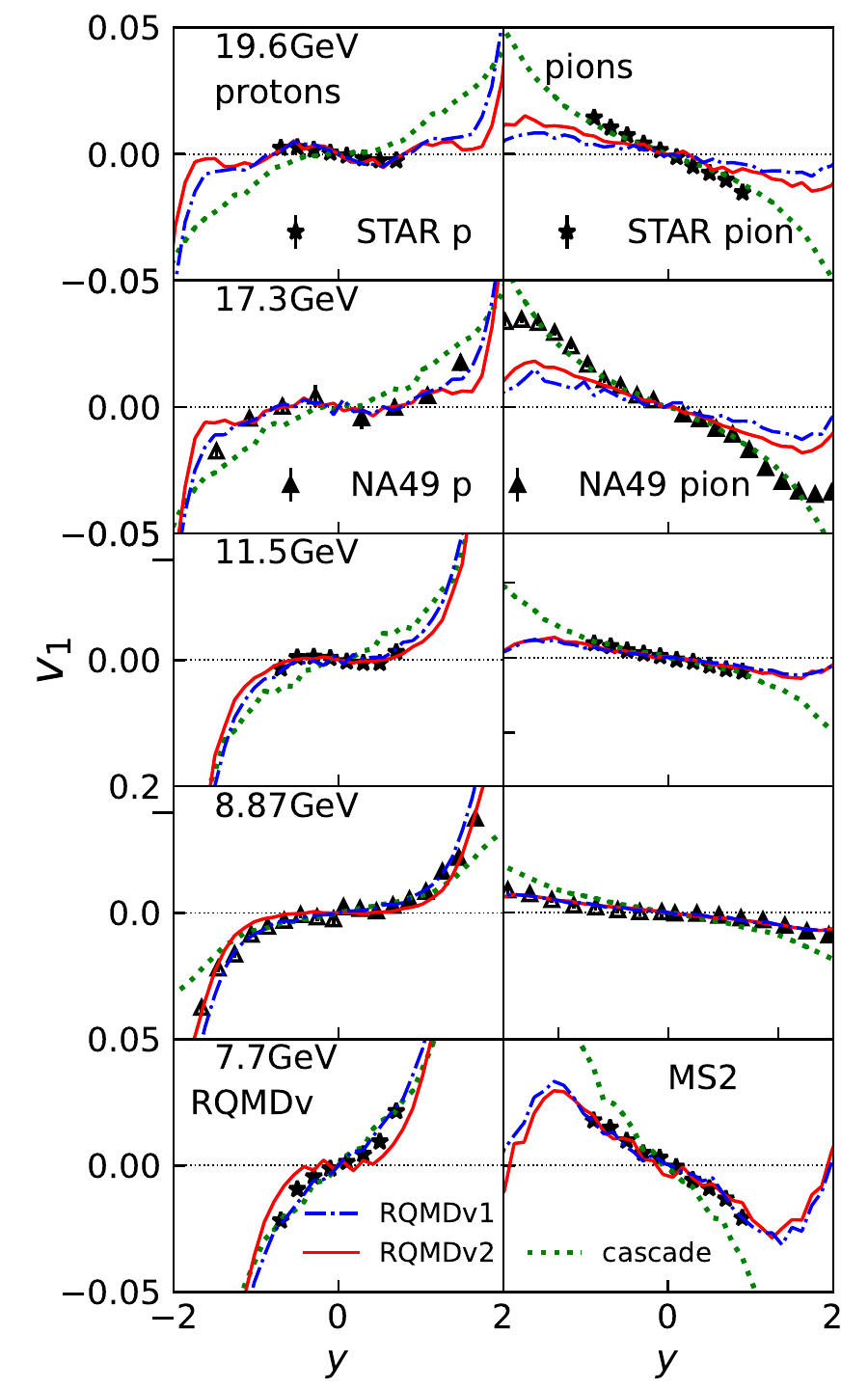}
\caption{Rapidity dependence of the directed flow $v_1$ in mid-central Au + Au collisions at
$\sqrt{s_{NN}}=7.7, 11.5, 19.6$ GeV 
and mid-central Pb + Pb collisions at $\sqrt{s_{NN}}=8.87$ and 17.3 GeV in comparison with the NA49~\cite{NA49prc} and STAR~\cite{STARv1} data.
In the left panel, we show the results
from cascade mode (dotted lines), RQMD/S (dotted-dashed lines),
RQMDs1 (dashed lines), and RQMDs2 (solid lines).
The left panels show the proton $v_1$ and the right panels for pion $v_1$.
In the right panel, we show the results from 
cascade mode (dotted lines),
the RQMDv1 (dashed-dotted lines)
and RQMDv2 (solid lines) model prediction with the MS2 EoS. 
}
\label{fig:v1RQMD}
\end{figure*}

In the left panel of Fig.~\ref{fig:v1RQMD}, we compare the rapidity dependence of the $v_1$
of protons and pions
from cascade mode, RQMD/S, RQMDs1, and RQMDs2
with the STAR~\cite{STARv1} and NA49 data~\cite{NA49prc}.
We select the impact parameter range $4.6 < b< 9.4$ fm to compare
10-40\% central Au + Au collisions at 7.7, 11.5 and 19.6 GeV,
and Pb + Pb collisions at 8.87 and 17.3 GeV.
It is seen that the RQMD/S results are in good agreement with the RQMDs1 results for all incident energies, while RQMDs2 shows somewhat less $v_1$ than the other models.
All models predict negative proton $v_1$ slope above $\sqrt{s_{NN}}=10$ GeV except for cascade results.
The negative $v_1$ is mainly generated during the expansion stage after
two nuclei pass through each other. Additional potential interaction generates
more negative flow.
However, when we use stronger potential by taking smaller width $L=0.5$ fm$^2$,
three models, RQMD/S, RQMDs1, and RQMDs2,
predict positive $v_1$ for protons at 11.5 GeV, which demonstrates
the strong sensitivity of the slope of the proton directed
flow to the interaction; both weak and strong interaction generate positive proton flow.
We will discuss the dynamical origin of a negative flow in our model later.

\begin{figure*}[tbh]
\includegraphics[width=8.0cm]{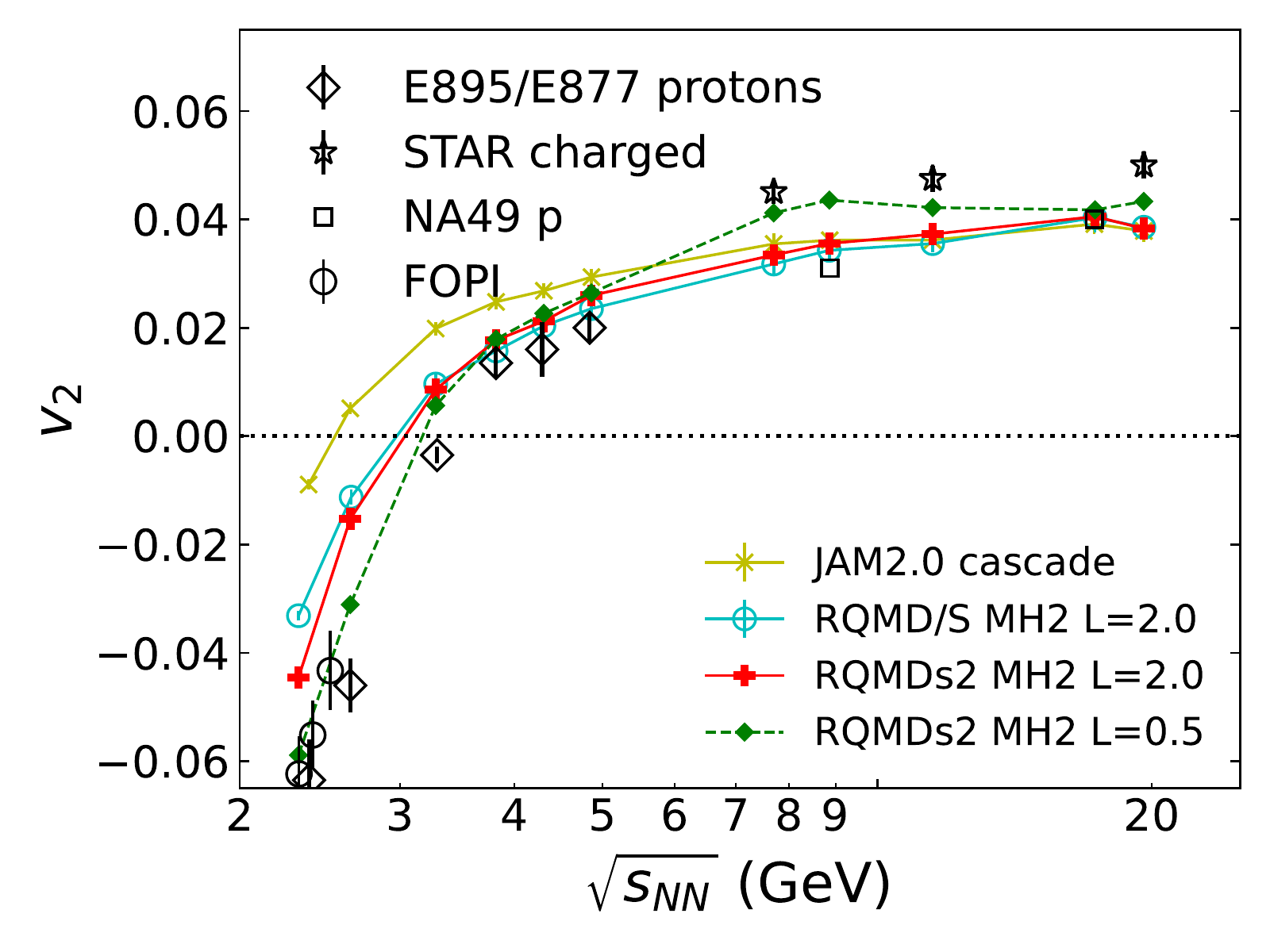}%
\includegraphics[width=8.0cm]{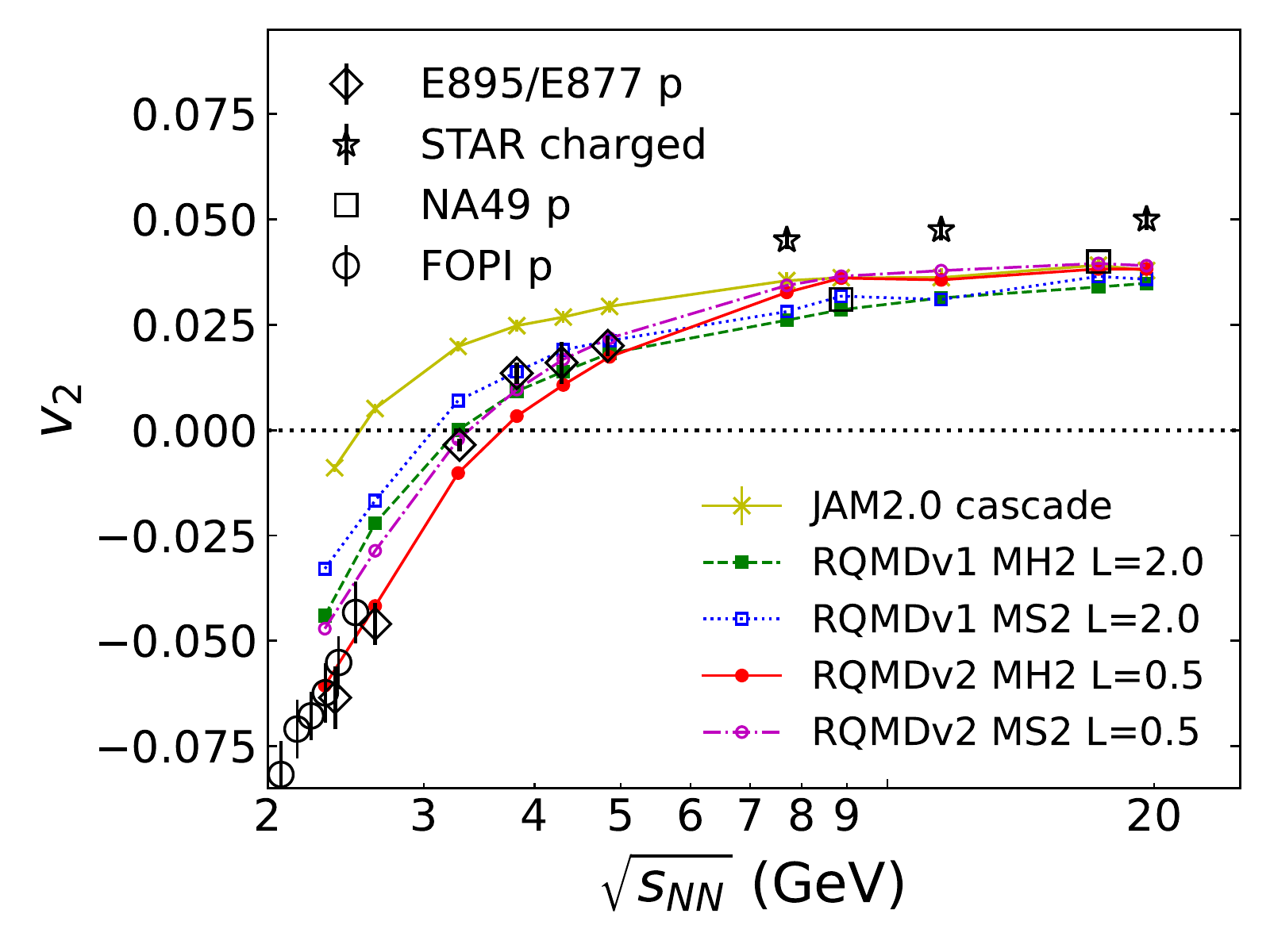}
\caption{The beam energy dependence of the proton elliptic flow
at mid-rapidity are compared with the data from FOPI~\cite{Andronic:2004cp},
E895/E877~\cite{E895v2},
NA49~\cite{NA49prc},
and STAR~\cite{STARv2}.
In the left panel, we show the results
from cascade, RQMD/S and RQMDs2 
with the hard momentum-dependent EoS (MH2).
In the right panel,
we show the cascade, RQMDv1 and RQMDv2 results.
}
\label{fig:v2RQMD}
\end{figure*}

In the left panel of Fig.~\ref{fig:v2RQMD},
the beam energy dependence of the proton elliptic flow $v_2$
at mid-rapidity in mid-central Au + Au collisions
from the cascade mode, RQMD/S and RQMDs2 are compared with the data.
RQMDs1 results are not plotted because they are almost identical to the RQMDs2 results.
Elliptic flow is also consistent among the models.
The calculations with $L=2.0$ fm$^2$ predict less squeeze-out compared with
the data below $\sqrt{s_{NN}}<5$ GeV.
We found that RQMSs2 with $L=0.5$ fm$^2$ improves the description of the
elliptic flow.

In this section, we have demonstrated that
two different implementations of
the Skyrme force as a Lorentz scalar 
in the quantum molecular dynamics
approach: RQMD/S and RQMDs
yield almost the same results
at relativistic energies, and they improve the description of the data
from the cascade model simulations.
However, the scalar potential does not generate enough pressure to reproduce
anisotropic collective flows at AGS energies.
In the next section, we will  discuss results from the
RQMDv model, in which the Skyrme potential 
is incorporated as a Lorentz vector potential.

\subsection{RQMDv: RQMD with Lorentz vector potential}

We shall now present the results of the RQMD model, in which
the Skyrme potential is implemented as a Lorentz vector (RQMDv).
Similarly to the previous section,
the RQMDv1 model refers to the model which uses the two-body distance Eq.~(\ref{eq:twobodydistance})
for the argument of potential,
while RQMDv2 uses Eq.~(\ref{eq:restframedistance}).

The right panel of Fig.~\ref{fig:pxRQMD} shows the sideward flow from the RQMDv2 model
for mid-central Au + Au collisions at $\sqrt{s_{NN}}=2.7, 3.3, 3.8$ and 4.3
GeV. It is seen that vector potential predicts stronger flow
than the scalar potential, and a good description of the data is
obtained for the soft momentum-dependent EoS from MS2 for both Gaussian
widths of $L=0.5$ and 2.0 fm$^2$ and hard momentum-dependent EoS from MH2 with $L=2.0$ fm$^2$.
The hard momentum-dependent MH2 EoS with $L=0.5$ fm$^2$ overestimates the data.
We note that the RQMDv1 results are almost identical to those of RQMDv2 results
at AGS energies.
We also note that the parameters MH2 are not as hard as the hard EoS at high
baryon densities, as the parameter which controls the repulsive part of the 
potential is $\gamma=1.67$ in MH2, while  it is $\gamma=2$ in the hard
momentum-independent EoS. 
If we compare the parameter sets MH1 and MS1, we have
slightly stronger EoS dependence than the parameter sets of MH2 and MS2.

The right panel of Fig.~\ref{fig:v1RQMD} shows the directed flow from the RQMDv1 model with $L=2.0$ fm$^2$ 
and RQMDv2 with $L=0.5$ fm$^2$
for protons (left panels) and pions (right panels).
The EoS from MS2 is used in the calculations.
Both the RQMDv1 and RQMDv2 models describe the beam energy dependence of
the proton directed flow data at SPS energies.
Our models correctly predict the negative pion directed flow for all beam
energies, which is due to the shadowing effects by the participant matter.
As beam energy is increased, the models predict less slope than the data.
We may need to include the pion potential.

\begin{figure}[tbh]
\includegraphics[width=8.5cm]{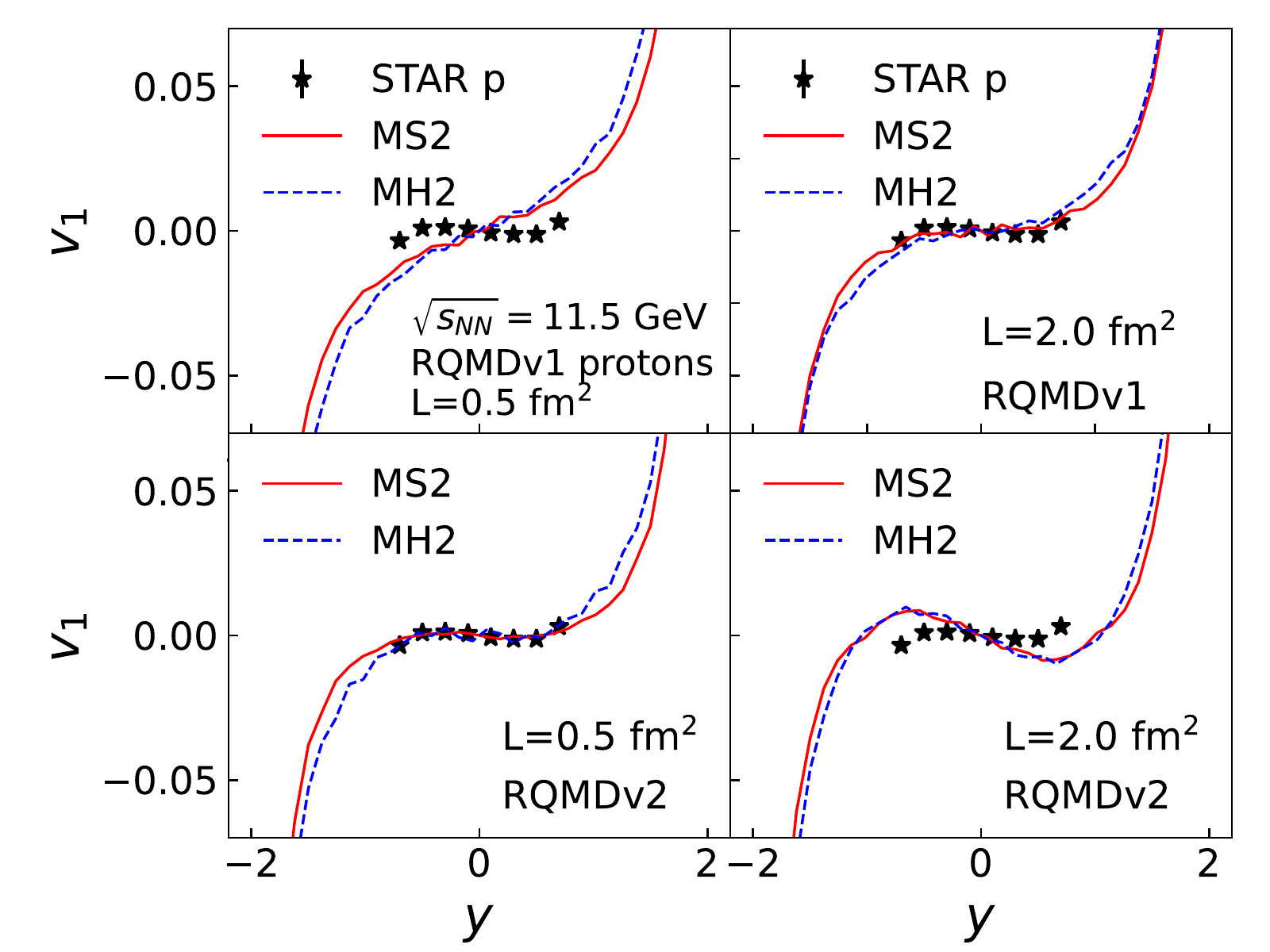}
\caption{EoS and the Gaussian width dependence on the directred flow in mid-central Au + Au collision at $\sqrt{s_{NN}}=11.5$ GeV
from RQMDv1 (upper panel) and RQMDv2 (lower panel)are
compared with the STAR data.
}
\label{fig:v1_11RQMDv}
\end{figure}

In Fig.~\ref{fig:v1_11RQMDv}, we examine the
EoS and the Gaussian width dependence on the directed flow
for mid-central Au + Au collision at 11.5 GeV.
We found that the $v_1$ slope becomes positive for $L=0.5$ fm$^2$ in RQMDv1,
while it is negative in RQMDv2. On the other hand, the $v_1$ slope
from both RQMDv1 (with MS2) and RQMDv2 is negative for $L=2.0$ fm$^2$.
We note that $L$ controls the strength and the range of the potential
interaction in the QMD type model.
These results indicate that
the sign of the $v_1$ slope is highly
sensitivity to the strength of the interaction.

In Fig.~\ref{fig:v1_11RQMDv}, hard and soft momentum-dependent EoS are compared.
It is seen that $v_1$ is not sensitive to the EoS at 11.5 GeV, 
which is understood by the fact that
Au + Au collisions at 11.5 GeV do not provide high baryon densities
due to the partial stopping of the nuclei.

The right panel of Fig.~\ref{fig:v2RQMD} compares the elliptic flow of protons
at mid-rapidity from the RQMDv1 and RQMDv2 models
for the mid-central Au + Au and Pb + Pb collisions
with the experimental data.
The EoS dependence between momentum-dependent soft and hard EoS
for the elliptic flow is seen for the beam energies below 5 GeV.
The effect of the Gaussian width on the elliptic flow is also seen, which
indicates the sensitivity to the interaction strength. 
The elliptic flow  using the smaller width of $L=0.5$ fm$^2$ 
is slightly larger above 7 GeV due to a more negligible shadowing effect
because of the shorter interaction range than in the calculations
with $L=2.0$ fm$^2$.
On the other hand, a shorter width predicts a stronger shadowing effect
at energies less than 5 GeV due to stronger interaction.

\section{inspections of collision dynamics}

We now investigate the collision dynamics of how the directed flow is generated
within our model. 
First, let us summarize the role of the collision term
for the generation of the directed flow.

In Ref.~\cite{Zhang:2018wlk}, we investigate the effects of spectator and
meson-baryon interactions on the flow within a cascade model.
When secondary interactions (mainly meson-baryon and meson-meson collisions) are
disabled, negative proton flow is generated by the shadowing from the spectator 
matter at $\sqrt{s_{NN}}<30$ GeV, while above 30GeV there is no shadowing
effect, and directed flow is not generated by the initial Glauber-type
nucleon-nucleon collisions.
The effect of the secondary interaction is to generate positive directed flow
at $\sqrt{s_{NN}}<30$ GeV since secondary interactions can start
\textit{before} two nuclei pass through each other. In contrast, negative flow is generated
by the secondary interactions at $\sqrt{s_{NN}}>30$ GeV,
since secondary interactions start \textit{after} two nuclei pass through each other due to a tilted expansion.
We note that tilted expansion of the matter created in noncentral heavy-ion collisions is a general feature of the dynamics
for a wide range of collision energies.

\begin{figure}[tbh]
\includegraphics[width=8.9cm]{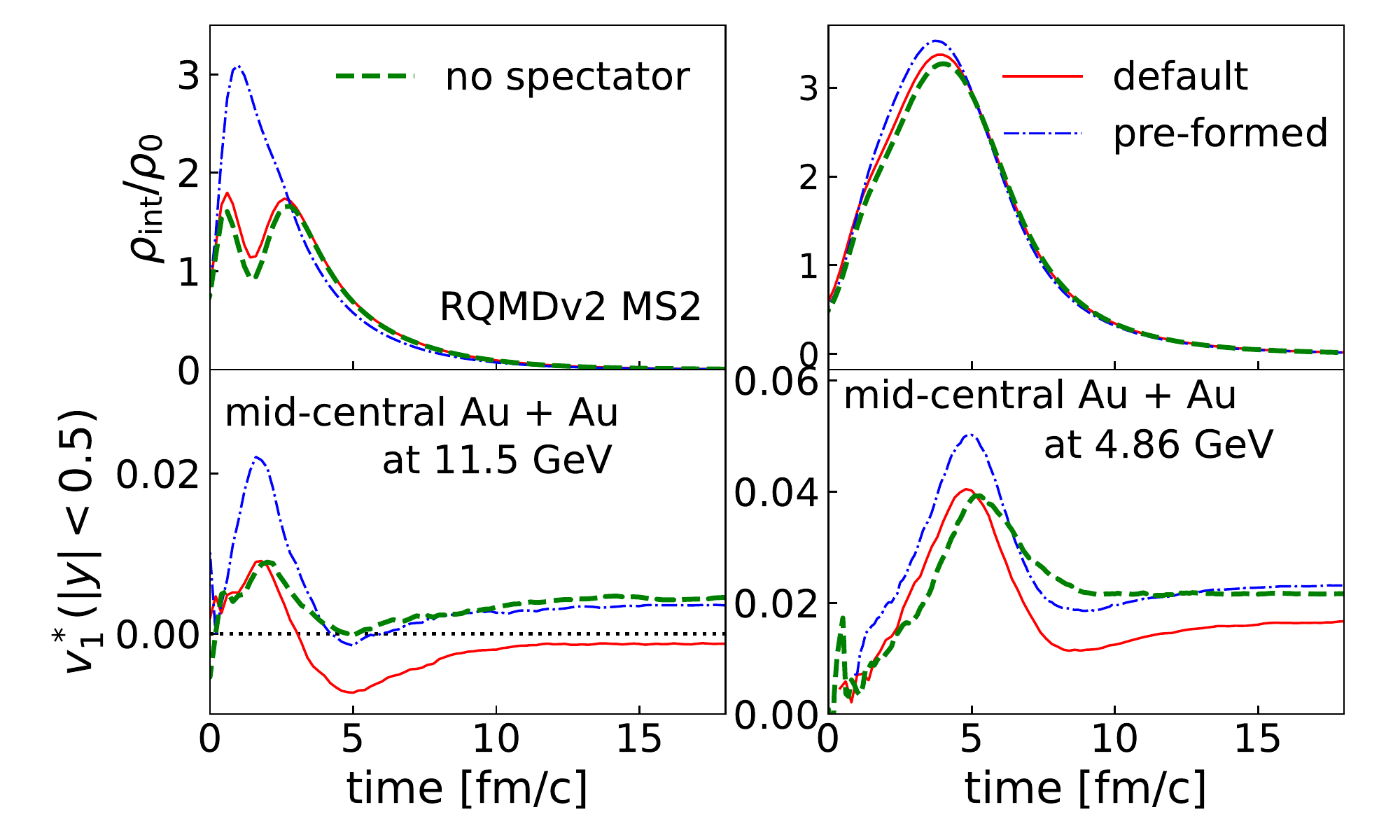}
\caption{Time evolution of the invariant interaction density (upper panel)
averaged over the central cell of  $|x|\leq3$ fm, $|y|\leq3$ fm, and $|z|\leq1$ fm and sign weighted directed flow $v_1^*$ of baryons at mid-rapidity $|y|<0.5$
(lower panel)
for mid-central Au + Au collisions at $\sqrt{s_{NN}}=11.5$ GeV
from the RQMDv2 calculation are shown in the left panels.
Right panels show the same but for the beam energy of 4.86 GeV.
The solid lines show the results from default calculations.
The dotted-dashed lines show the results of the calculations
that include the potential interaction for pre-formed baryons.
The dashed lines represent the results of the calculation without
interactions of spectator matter.
}
\label{fig:v1evolution}
\end{figure}
Let us now investigate the time evolution of the directed flow.
In Fig.~\ref{fig:v1evolution}, we show in the upper panel the time evolution of 
the invariant interaction density $\rho_{B,i}=\sqrt{J_i^2}$,
which is used for the density dependent part of the potentials
in Eq.~(\ref{eq:vectorpotential})
for mid-central Au + Au collisions at 11.5 GeV (left panel) and 4.86 GeV (right panel).
In the lower panel of Fig.~\ref{fig:v1evolution},
we plot the sign weighted directed flow for baryons integrated
over a rapidity range of
$|y|<0.5$:
\begin{equation}
 v_1^* = \int^{0.5}_{-0.5} dy\, v_1(y)\mathrm{sign}(y).
\end{equation}
A general feature of the temporal evolution of the directed flow at mid-rapidity
is that it rises within a first few fm/$c$ during the compression stages of
the reaction and then decreases in the expansion stages for both 11.5 and 4.86 GeV.
Finally, the flow goes up slowly at the very late stages of the collisions.
Only formed baryons feel the potentials in the default simulation (solid lines), 
although the constituent quarks can scatter in the
prehadrons.
At 11.5 GeV beam energy, most of the nucleons are excited to strings,
which results in the dip in the interaction density evolution at early times.

To see the effects of a possible potential interaction for the preformed
baryons, we include the potential interaction for the preformed leading
baryons that have original constituent quarks with the reduced factor of 1/3 (one quark)
or 2/3 (diquark)~\cite{Li:2007yd},
which is shown by the dotted-dashed lines.
Additional potential interaction generates
two times more positive directed flow in the compression stage of the collision.
It is also worthwhile to recognize that the directed flow decreases quickly
even for the stronger interaction by generating more negative directed flow
in the expansion stages.
We should emphasize that density-dependent interactions (hard or soft)
do not predict negative directed flow in our framework because the interaction is
weak, and it generates a small amount of negative flow
during the expansion stage.
When only density-dependent potentials are included,
the potential becomes attractive in the expansion stages at 11.5 GeV
as the baryon density is around the normal nuclear density, 
which prevents developing the flow, and
positive flow developed in the compression stage remains positive at freeze-out.
In contrast, in the case of momentum-dependent potential,
the attractive force is mainly
generated by the momentum-dependent part of the potential,
and the density-dependent part is repulsive. 
In the expansion stages, momenta of particles are random,
and the momentum-dependent part of the potential becomes weak:
as a result, the net effect of the interaction is repulsive,
which contributes to generating strong negative flow.
The negative proton directed flow at around 11.5 GeV beam energies
can only be obtained for an appropriate amount of the interaction strength:
weak interaction (including cascade mode) does not generate strong anti-flow,
on the other hand, strong interaction generates a very large positive flow at 
the compression stage.

To investigate the primary mechanism of decreasing directed flow in the expanding
stages of the collisions,
we have checked the effect of the spectator-participant interaction on the
directed flow by disabling the interaction between them.
``Spectator'' is defined as the nucleons which will not collide in the
sense of the Glauber type initial nucleon-nucleon collisions.
Specifically, we first compute the number of nucleon-nucleon collisions,
and we remove nucleons from the system if they will not interact with other nucleons.
It is seen that when ``spectators'' are not included in the simulation,
the directed flow decreases less in the expansion stages and the directed flow
remains positive at 11.5 GeV.
The shadowing effect by the spectator matter is large even at 11.5 GeV.

\begin{figure}[tbh]
\includegraphics[width=8.0cm]{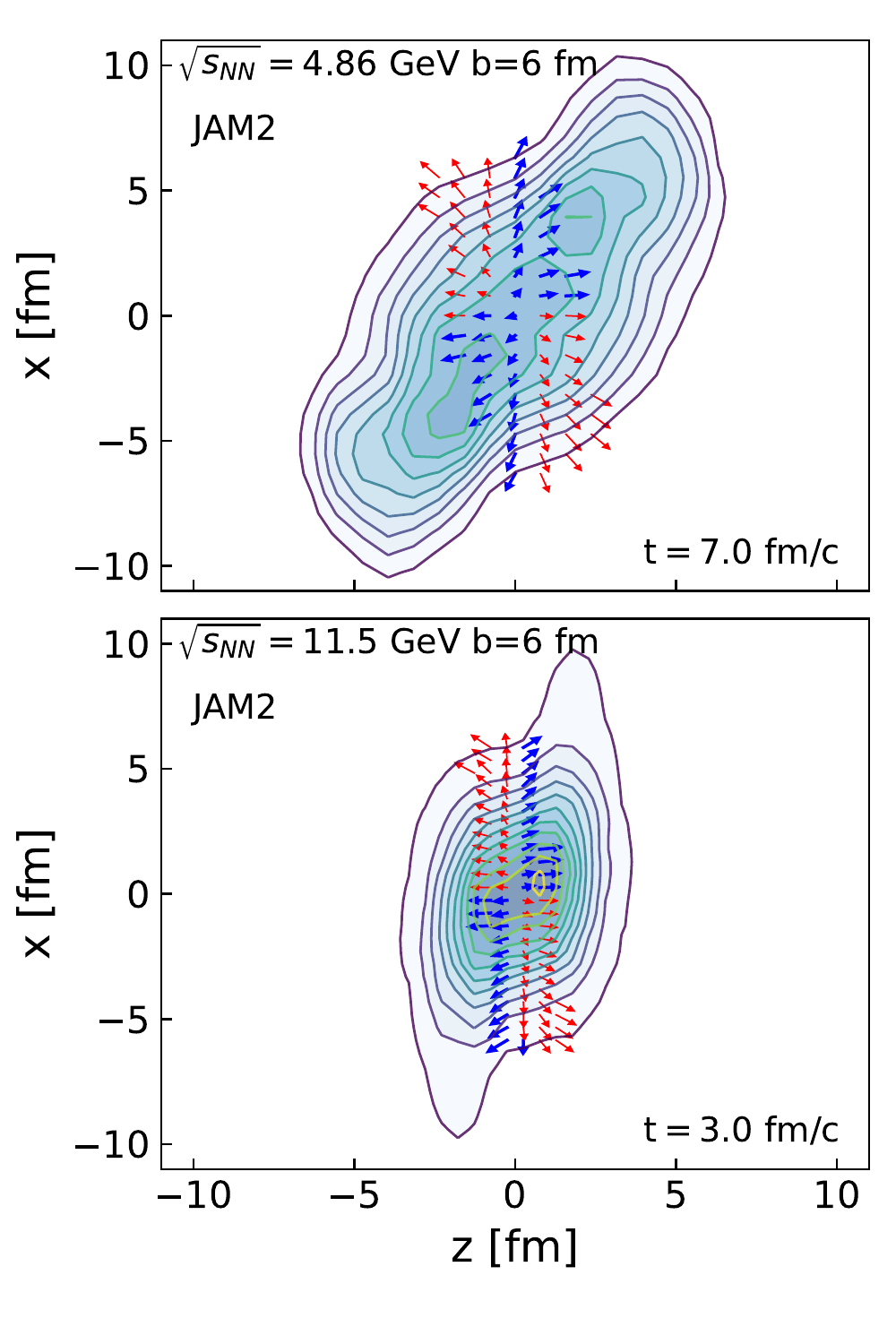}
\caption{Baryon density distribution at the time $t=7$ fm/$c$ in Au + Au collision
at $\sqrt{s_{NN}}=4.86$ GeV at the impact parameter $b=6$ fm (upper panel)
and $t=3$ fm/$c$ at $\sqrt{s_{NN}}=11.5$ GeV (lower panel).
The thin (red) arrows  show the local velocity of anti-flow,
and the bold (blue) arrows indicate normal flow for mid-rapidity $|y|<0.5$.
}
\label{fig:contour}
\end{figure}

Besides the shadowing effect,
the main dynamical origin of decreasing behavior of the directed flow at
mid-rapidity in the expansion stages of the collision
is the creation of an ellipsoid tilted with respect to the beam
axis, which generates antiflow predominately over the normal flow
as discussed in detail in Ref.~\cite{Brachmann:1999xt} within a 3FD model.
We argue that this mechanism is general; it holds true for all
high-energy noncentral heavy-ion collisions.
In Fig.~\ref{fig:contour}, we plot baryon density and local velocity for
mid-rapidity $|y|<0.5$ at $t=7$ fm/$c$
for Au + Au collisions at $\sqrt{s_{NN}}=4.86$ GeV at the impact parameter
$b=6$ fm (upper panel) and $t=3$ fm/$c$ at $\sqrt{s_{NN}}=11.5$ GeV (lower panel).
It is seen that a tilt of the matter distribution is created for both energies.
The tilted matter is a general consequence of the collision dynamics for
noncentral collisions, which is caused by the initial nucleon-nucleon collisions and the degree of nuclear stopping.
 
At lower energies, the compression time is long enough to create a large directed flow, mainly due to a strong repulsive interaction.
At late times, when the system
starts to expand, antflow wins against normal flow. However, interaction
is weaker during the expansion as compared to compression stages at lower
energies. At very late times, after the tilted matter is smeared out,
the flow turns to go up again.
The net effect is to create a positive directed flow at freeze-out.
When going to higher beam energies, compression time becomes shorter, so less positive flow 
is generated. Because expansion time is longer than the compression time at
high energies, more negative flow is generated.
In addition, at high energies, many secondary particles are created,
and baryons can interact with mesons more, which results in the generation of
more negative flow.
This is the main mechanism of the beam energy dependence of the proton directed flow.
Therefore, the dependence of the transition from positive to negative flow may
reveal important information on the strength of the interactions of the excited matter.

\section{Results for hadronic spectra}
\label{sec:bulk_spectra}

In this section, we show the results of the JAM2 approach
for the bulk observables such as the rapidity distribution
and the transverse momentum spectra of protons and pions to
demonstrate the influence of the mean-field potentials on the bulk observables.

\begin{figure}[h]
\includegraphics[width=8.5cm]{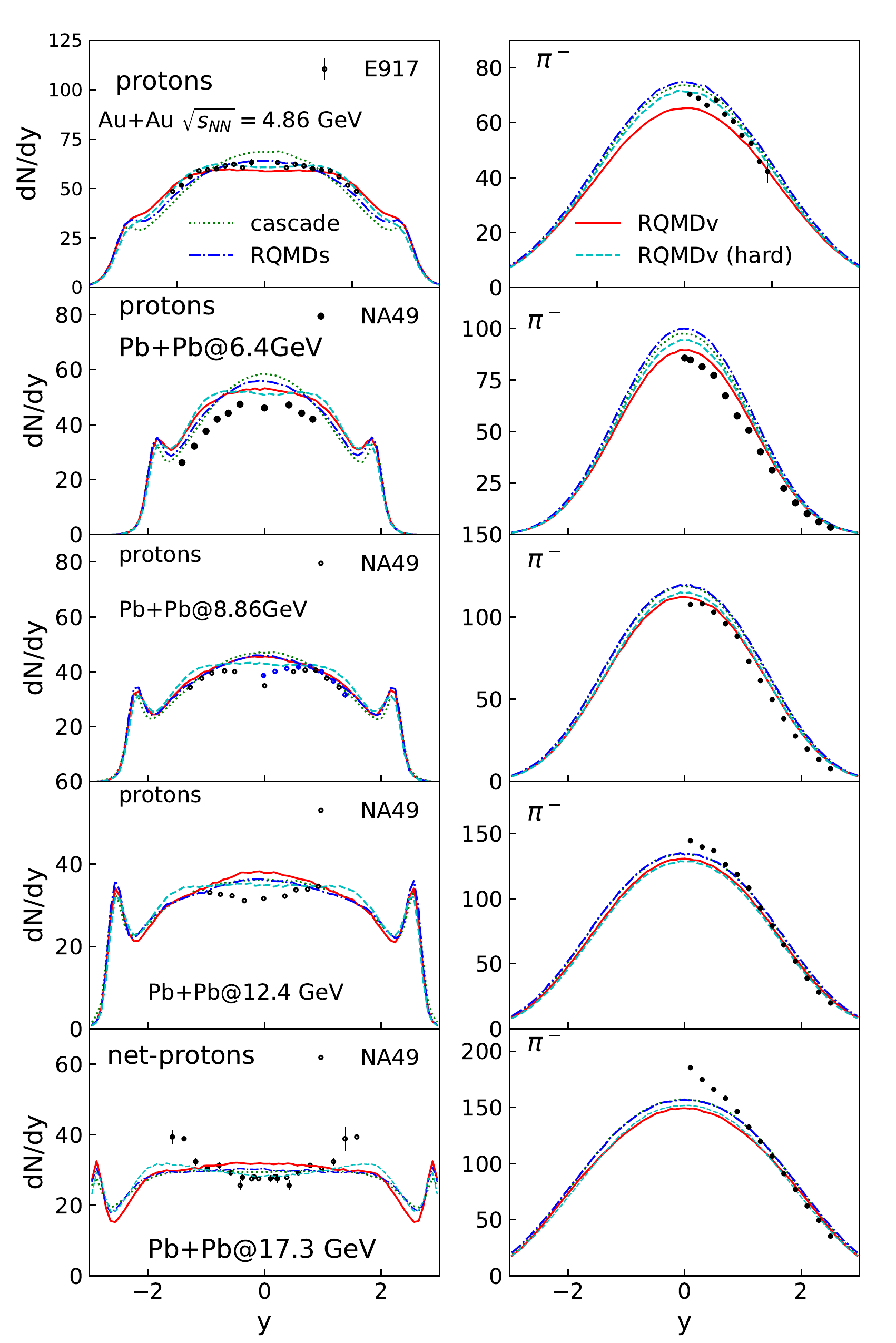}
\caption{The rapidity distributions of protons (left panels) and negative
pions (right panels)
in central Au + Au collisions at 4.86 GeV and Pb + Pb collisions 
at $\sqrt{s_{NN}}=6.4$, 8.86, 12.4, and 17.3 GeV.
In the left panel, we show the results 
from the cascade mode (dotted lines), 
RQMDs with MS2 (dotted-dashed lines), RQMDv with MS2 (solid lines),
and RQMDv with hard EoS (dashed lines).
The experimental data have been taken from 
Refs.~\cite{e802,e917,NA49:2002pzu,Blume:2007kw,NA49:2007stj,NA49:2010lhg}.
}
\label{fig:dndyRQMD}
\end{figure}

The EoS and Gaussian width $L$ dependence are less sensitive to the rapidity
and transverse momentum distributions than anisotropic flows.
We present the results for MH2 EoS and $L=2.0$ fm$^2$
for RQMDs and $L=0.5$ fm$^2$ for RQMDv calculations.
The systematic study will be presented elsewhere.

In Fig.~\ref{fig:dndyRQMD}, we show the proton and negative pion rapidity distributions
in central Au + Au collisions at $\sqrt{s_{NN}}=4.86$ GeV
and central Pb + Pb collisions
at $\sqrt{s_{NN}}=6.4$, 8.86, 12.4, and 17.3 GeV.
The results from cascade, RQMDs, and RQMDv models are compared with the experimental data
from E802~\cite{e802}, E917~\cite{e917}, and NA49~\cite{NA49:2002pzu,Blume:2007kw,NA49:2007stj,NA49:2010lhg} Collaborations.
The RQMDs results are almost the same as the cascade results due to the
weakness of the scalar potential, while the effects of the vector potential are visible. 
The influence of the mean-field potential on the proton rapidity distribution
is to reduce the stopping of protons except for the RQMDv calculations
with momentum-dependent potential at high energy $\sqrt{s_{NN}}>10$ GeV.
The RQMDv model with momentum-dependent potential predicts slightly more stopping at higher beam energies due to the disappearance of the attractive momentum-dependent force.
To see the effects of momentum-dependent potential, we also plot the results
from the RQMDv (hard) calculation with the momentum-independent hard EoS.
The RQMDv (hard) model predicts less stopping compared with the cascade calculations.
It is seen that the mean-field potential reduces the pion multiplicity
in the case of the vector potential.

\begin{figure}[hbt]
\includegraphics[width=8.5cm]{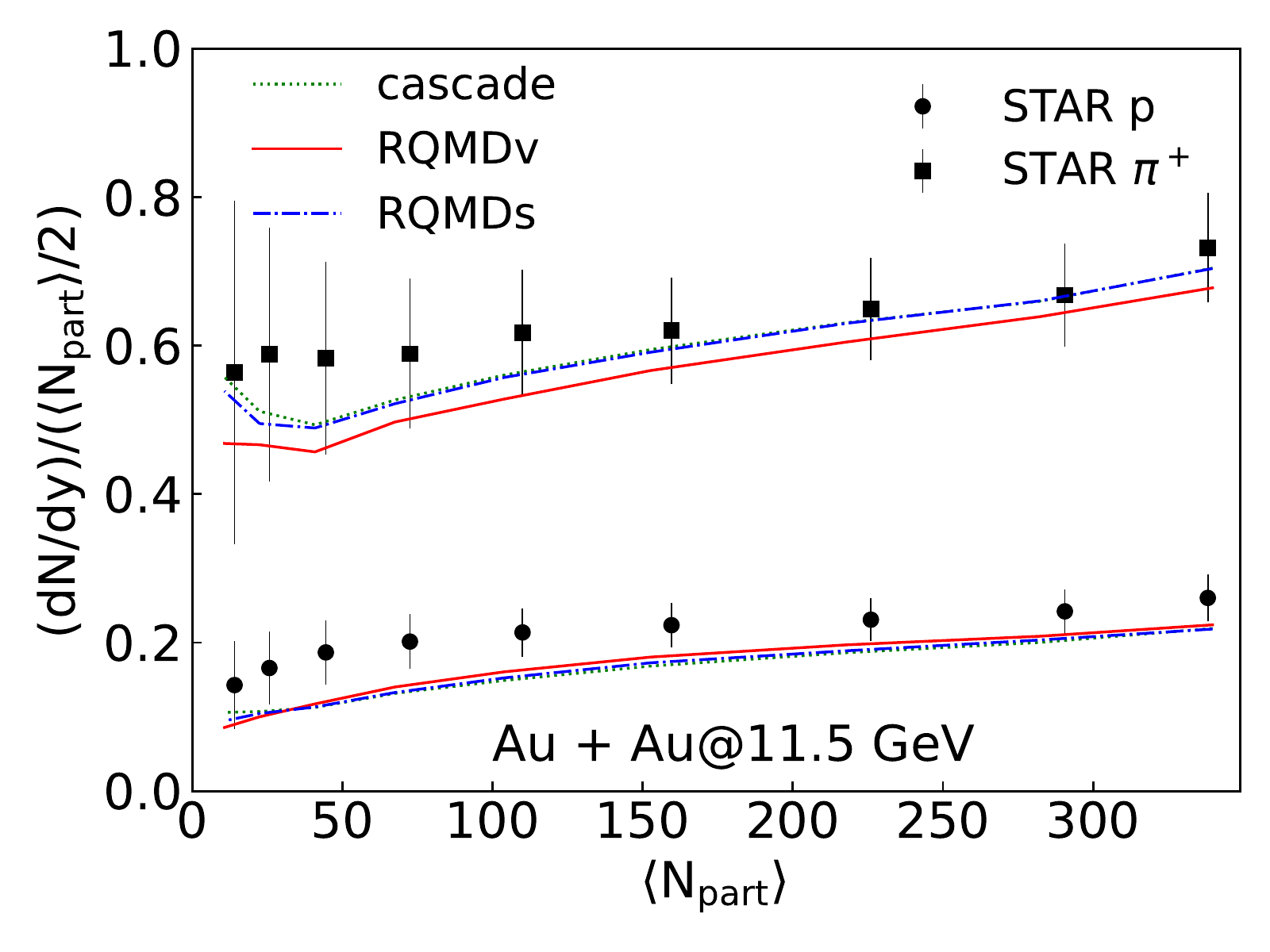}%
\caption{The centrality dependence of 
$dN/dy/(\langle N_\mathrm{part}\rangle/2)$
for protons and negative pions in Au + Au collisions at 11.5 GeV.
The cascade (dotted lines), RQMDs (dotted-dashed lines), and RQMDv (solid lines) results
are compared with the STAR data~\cite{STAR:2017sal}.
}
\label{fig:npartRQMD}
\end{figure}

In Fig.~\ref{fig:npartRQMD}, we compare the centrality dependence of
the proton and positive pion $dN/dy$ normalized by  
the mean value of the number of participating nucleons $\langle N_\mathrm{part}\rangle$ for $|y|<0.1$
in Au + Au collisions at $\sqrt{s_{NN}}=11.5$ GeV
from cascade, RQMDs, and RQMDv with the experimental data from the STAR Collaboration~\cite{STAR:2017sal}.
The number of participants $\langle N_\mathrm{part}\rangle$
is obtained by counting the predicted initial nucleon-nucleon collisions before
nuclei collide, and extracting predicted participating nucleons.
This procedure is the same
as the Monte Carlo Glauber calculation~\cite{Miller:2007ri}.
The models reproduce the centrality dependence of both protons and pions,
while the proton multiplicity is slightly underestimated.

\begin{figure}[htb]
\includegraphics[width=8.5cm]{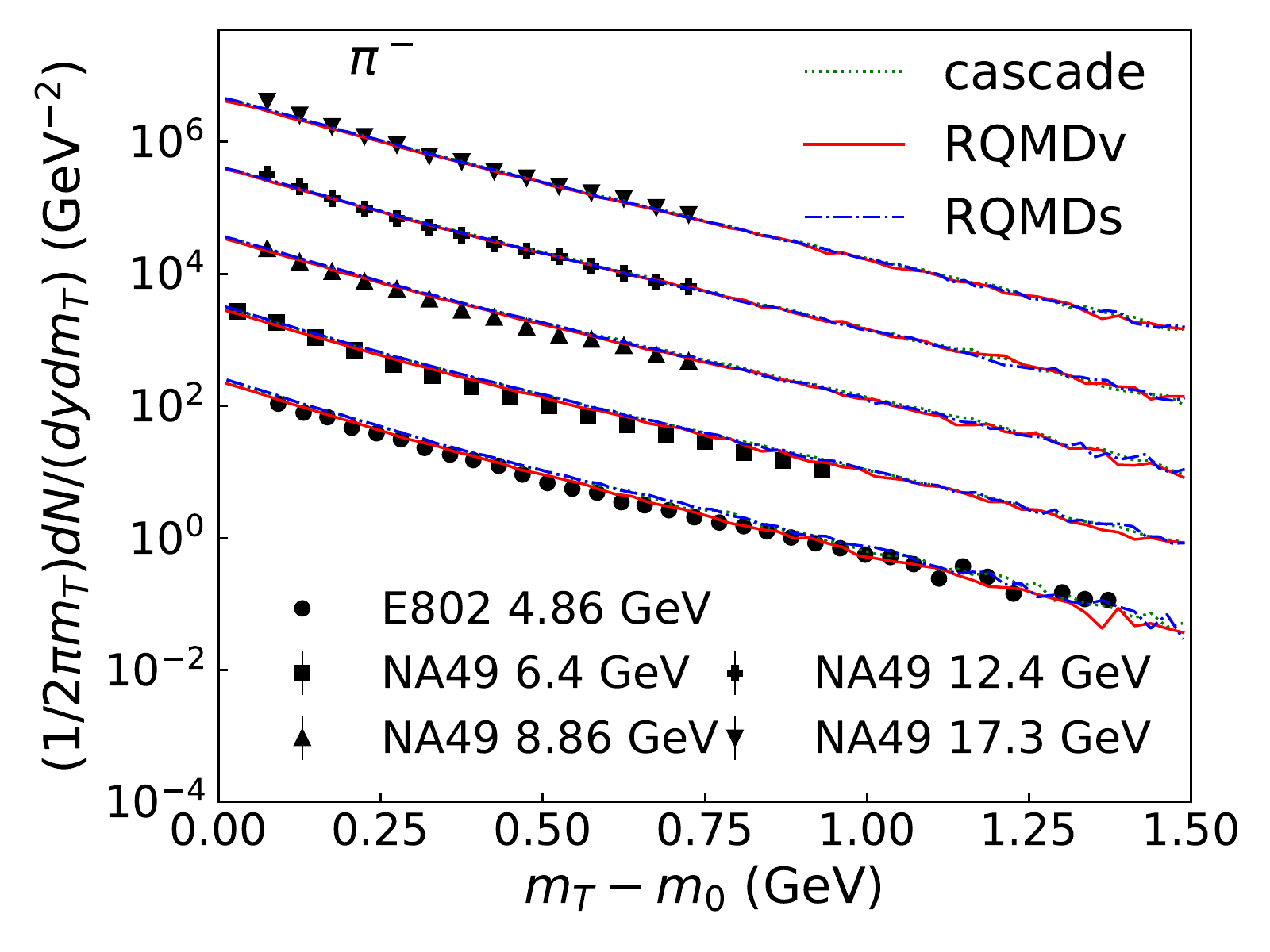}
\includegraphics[width=8.5cm]{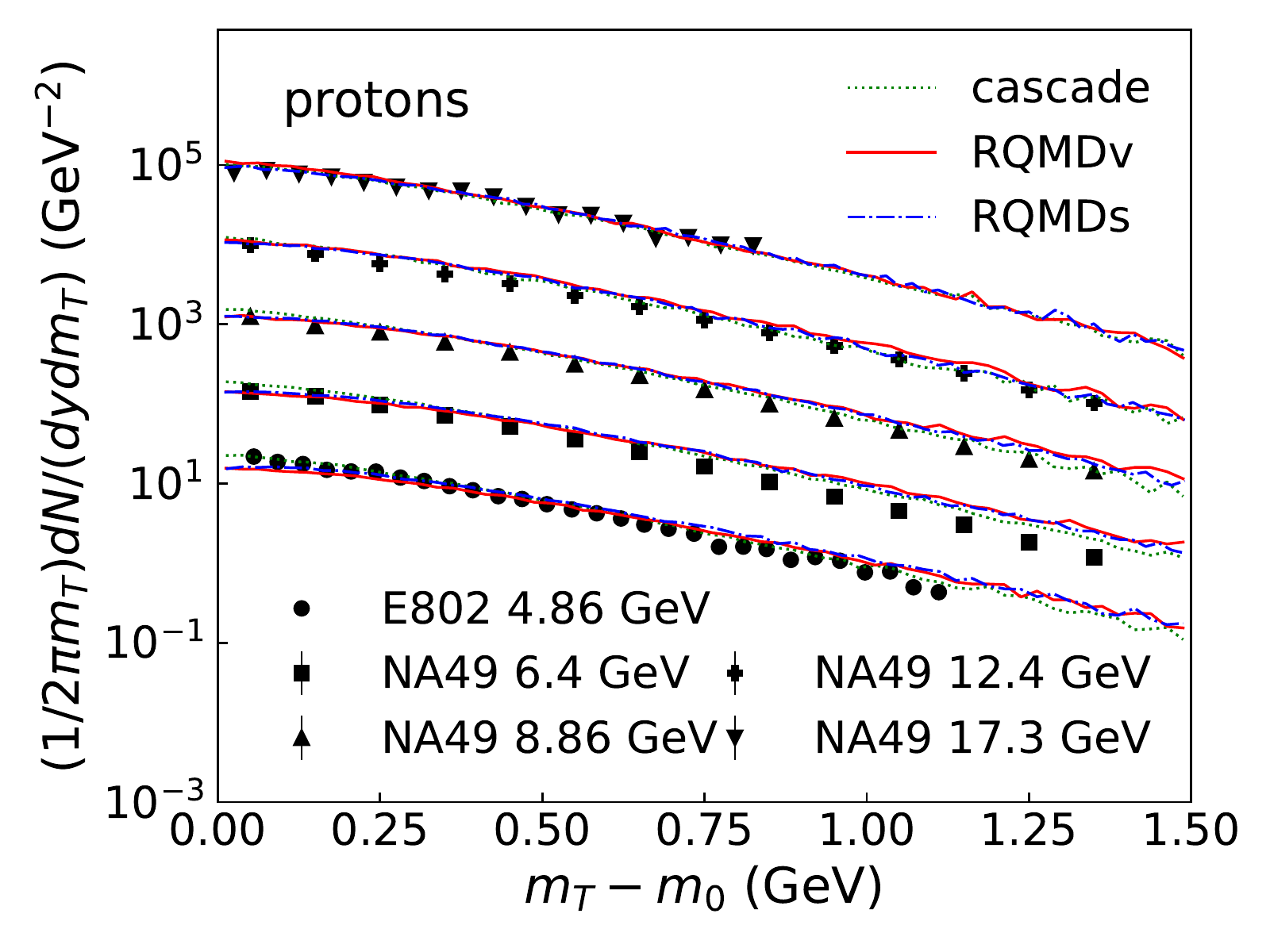}
\caption{The transverse mass spectra 
for negative pions (upper panel) and protons (lower panel)
are compared with central Au + Au collisions at $\sqrt{s_{NN}}=4.86$ GeV
and central Pb + Pb collisions at 6.4, 8.86, 12.4, and 17.3 GeV
from the E802~\cite{e802} and NA49 data~\cite{NA49:2002pzu,NA49:2007stj,NA49:2006gaj}.
The spectra other than 4.86 GeV data are increased by a factor of 10 from bottom to top.
}
\label{fig:dndmtRQMD}
\end{figure}
The proton and negative pion transverse mass $m_T=\sqrt{m_0^2+\bm{p}^2}$ spectra
of cascade, RQMDs, and RQMSv models in central Au + Au collisions at $\sqrt{s_{NN}}=4.86$ GeV and central Pb + Pb collisions at $\sqrt{s_{NN}}=6.4$, 8.86, 12.4, and 17.3 GeV
are displayed in Fig.~\ref{fig:dndmtRQMD}.
The model results are compared with the experimental data from the E802~\cite{e802}
and NA49 Collaborations~\cite{NA49:2002pzu,NA49:2007stj,NA49:2006gaj}.
The model predictions for both proton and pion transverse mass spectra
show a reasonable agreement with experimental data.
The effects of the potential are very small
on the transverse mass spectra,
except for the small suppression in the lower momentum region
reflecting the lower stopping of protons for $\sqrt{s_{NN}}<10$ GeV.

\begin{figure}[htb]
\includegraphics[width=8.5cm]{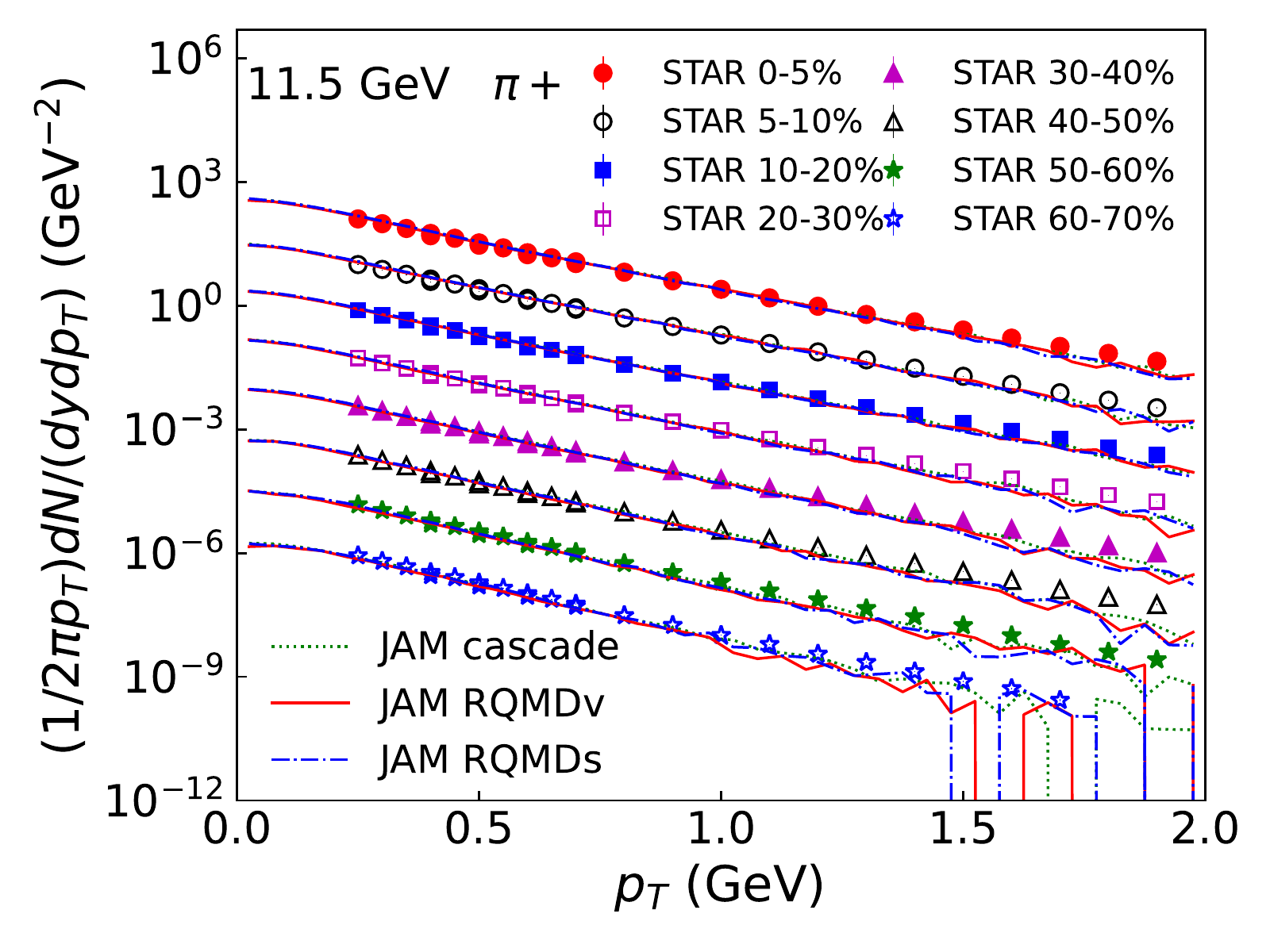}
\includegraphics[width=8.5cm]{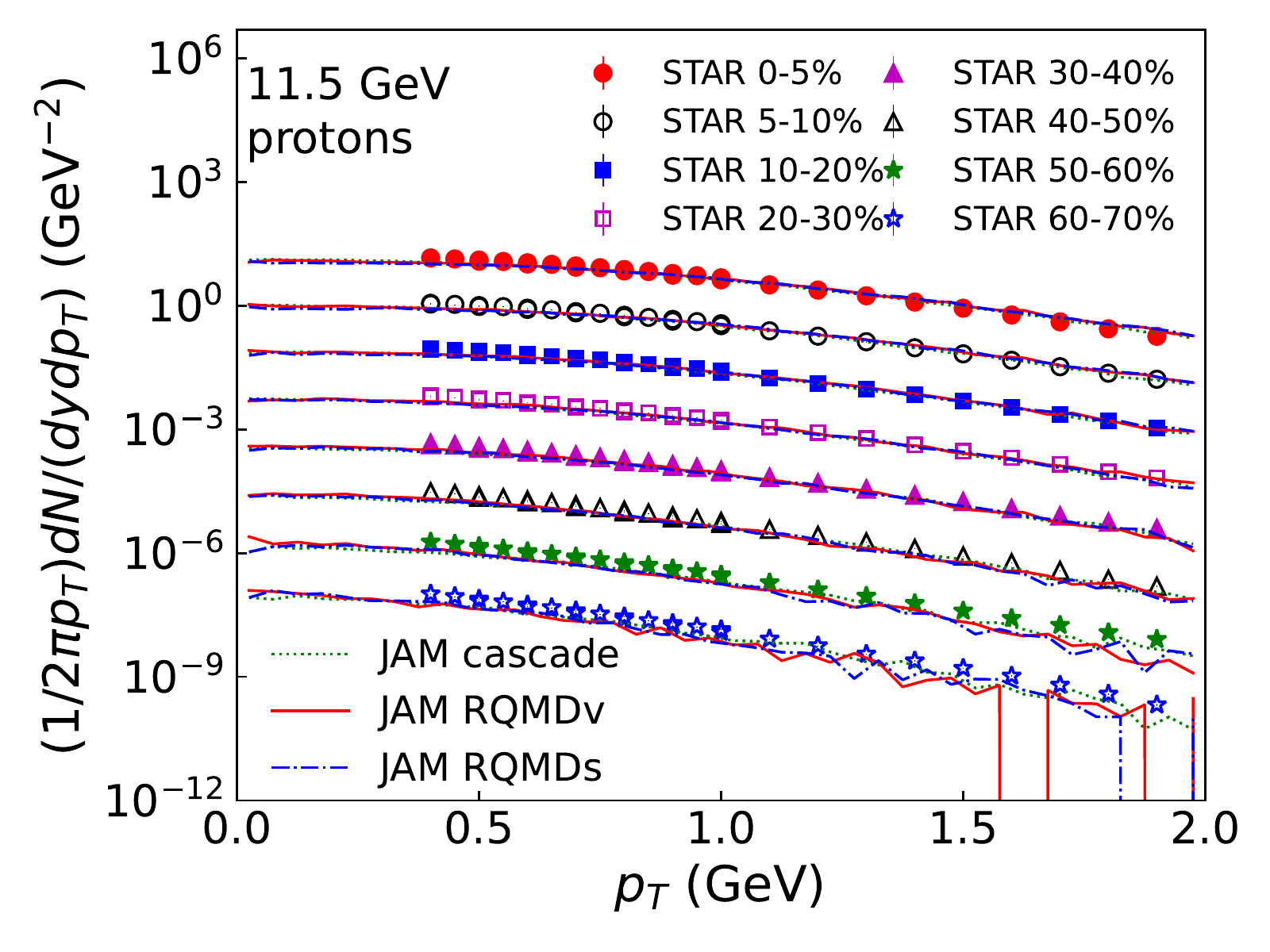}
\caption{The midrapidity ($|y|<0.1$) transverse momentum spectra 
for positive pions (upper panel) and protons (lower panel)
in Au + Au collisions at $\sqrt{s_{NN}}=11.5$ GeV
for different centralities.
The experimental data have been taken from Ref.~\cite{STAR:2017sal}.
The spectra other than 0-5\% central data are divided by a factor of 10 from top to bottom.
}
\label{fig:dndmtSTAR}
\end{figure}

We also examine the centrality dependence of the transverse momentum spectra
for protons and positive pions
in Au + Au collisions at $\sqrt{s_{NN}}=11.5$ GeV  in Fig.~\ref{fig:dndmtSTAR}. 
The model results are compared with the experimental data from the STAR
Collaboration~\cite{STAR:2017sal}. Good agreement between the model results and the experimental data is obtained.
We also see that predictions of all three models,
cascade, RQMDs, and RQMDv
for the transverse momentum are almost identical for all centralities.

\section{conclusion}
\label{sec:conclusion}

We have developed a new mean-field model by implementing
the Skyerme type potential into the RQMD framework
as a Lorentz scalar (RQMDs) or vector (RQMDv) potential, including momentum dependence.
The RQMDs and RQMDv models are realized by the event generator JAM2 code.
We have studied the mean-field effects on the directed and elliptic flows
by using RQMDs and RMDv.
RQMDs does not generate enough pressure at AGS energies and fails to reproduce the flow data,
while the RQMDs model describes the directed flow at $\sqrt{s_{NN}}>10$ GeV where net baryon number at mid-rapidity starts to decrease,
the and scalar potential plays a role rather than the vector potential.
In contrast, the RQMDv approach,
in which the Skyrme type potential is implemented as a Lorentz vector, 
generates a strong pressure at AGS energies and describes the flow data very well.
RQMDv also predicts the correct sign change
of the proton directed flow.
The conventional hadronic mean field explains
the negative directed flow of protons within our approach.

The slope of the directed flow at freeze-out
is determined by the delicate cancellation of the positive and
negative flows in high-energy mid-central heavy-ion collisions.
We found that the positive directed flow develops more than
the negative flow in the compression stages of the collisions,
while the more negative flow is developed
during the expansion stage due to a tilted expansion and shadowing by the
spectator nucleon matter.
At lower collision energies,
a large positive flow can be developed by the strong
repulsion due to high baryon density and long compression time.
The strength of the interaction is weaker at the late expansion stage 
because of lower baryon density.
A net effect is to have a positive flow at lower energies.
On the other hand,
at higher energies, where most secondary interactions start after
two nuclei pass through each other, there is not enough time to develop
positive flow during the short compression time.
In contrast to lower energies,
expansion time becomes longer due to a large number of produced particles.
A net effect is to generate negative flow.

The results strongly depend on the model parameters; thus, we will
not rule out the possible softening scenario within the current study.
To confirm that our model correctly describes the collision dynamics,
we need to perform more systematic data comparisons, e.g.,
the flow of strangeness particles. This line of work is in progress.

In this work,
we examine the effect of scalar or vector potential separately,
and we see that negative proton directed flow can only be obtained
with the momentum-dependent potential.
It is important to study other approaches such as the relativistic
mean-field theory, in which both scalar and vector interactions are included.
Recently, a transport approach with a vector potential
that includes a first-order transition and a critical point
was formulated~\cite{Sorensen:2020ygf}.
It will be interesting to use this potential to see the effects of
a phase transition on the flows.

Finally,
the inclusion of the mean fields into the hybrid model is an important
future work toward the complete description of
the space-time evolution of the system
in the high baryon density region.

\begin{acknowledgments}
We thank J. Steinheimer for valuable comments on the manuscript.
This work was supported in part by
Grants-in-Aid for Scientific Research from JSPS
(No. JP21K03577,
No. JP19H01898, 
and No. JP21H00121
).
\end{acknowledgments}

\appendix

\section{The correction of RQMD/S in 2005}
\label{sec:correction}

In this section, we correct the results of Ref.~\cite{Isse:2005nk},
and compare the corrected results for Au + Au collisions at $E_\mathrm{lab}=1.85A$ GeV.

The equations of motion Eq.(\ref{eq:motion1}) may be evaluated as
\begin{align}
\dot{\bm{r}_i} &= \frac{\bm{p}_i}{p^0_i}
    + \sum_{j\ne i}^N
     D_{ij}\frac{\partial q_{T,ij}^2}{\partial\bm{p}_i}
 + \sum_{j\neq i} E_{ij}\frac{\partial p_{T,ij}^2}{\partial\bm{p}_i},
      \\
\dot{\bm{p}_i} &= \  -\sum_{j\ne i}^N
  D_{ij}\frac{\partial q_{T,ij}^2}{\partial\bm{r}_i}.
\label{eq:motion2}
\end{align}
where 
\begin{align}
 D_{ij} &= \frac{\rho_{ij}}{4L} \left[
     \frac{m_i}{p_i^0}\frac{\partial V_{sk,i}}{\partial\rho_i} 
    +\frac{m_j}{p_j^0}\frac{\partial V_{sk,j}}{\partial\rho_j} 
   + \left(\frac{m_i}{p^0_i}+\frac{m_j}{p^0_j} \right) V_{m,ij} 
   \right], \label{eq:dij}\\
E_{ij} &= \rho_{ij}\left(\frac{m_i}{p^0_i}+\frac{m_j}{p^0_j} \right) V_{m,ij} 
       \frac{\partial V_{m,ij}}{\partial p_{T,ij}^2}. \label{eq:eij}
\end{align}
We note that Eqs.(A25) and (A26) in Ref.~\cite{Isse:2005nk},
which correspond to Eqs.~(\ref{eq:dij}) and (\ref{eq:eij}),
contain mistakes.
After correcting the mistake, we found that the results 
in Ref.~\cite{Isse:2005nk} were modified;
potential effects on the anisotropic flows become relatively smaller
in the RQMD/S approach. 
The results in Ref.~\cite{Nara:2015ivd} are also influenced by this mistake.
In this paper, we will present the corrected result.

The first factor in Eq.(A25) is wrong by a factor of 2:
 $\frac{1}{2L}$ should be $\frac{1}{4L}$.
 Furthermore, $1/(1-(p_{T,ij}/\mu_k)^2)$ in Eq.~(A26) 
 should be  $(1/(1-(p_{T,ij}/\mu_k)^2)^2$.
 In the code, only the first error was found.
We show the explicit expression here:
\begin{align}
\label{Eq:Dij}
 D_{ij}&=
\biggl(\frac{1}{4L}\biggr)\rho_{ij}
\left[
 \frac{\alpha}{2\rho_0}\biggl(\frac{m_i}{p_i^0}+\frac{m_j}{p_j^0}\biggr) \right.
 \nonumber\\
&\left.
+ \frac{\gamma}{1 + \gamma}
\frac{\beta}{\rho_0^\gamma}
\Biggl\{ \frac{m_i}{p_i^0}
\rho_i
^{\gamma -1}
\!\!+ \frac{m_j}{p_j^0}
\rho_j
^{\gamma -1}
\Biggr\}
\right]\nonumber \\
&+
\biggl(\frac{1}{4L}\biggr)
\frac{1}{2\rho_0}
\rho_{ij}
\biggl(\frac{m_i}{p_i^0}+\frac{m_j}{p_j^0}\biggr)
\sum_{k=1,2}
\frac{C_{k}}{1-[p_{T,ij}/\mu_k]^2},\\
\label{Eq:Eij}
E_{ij}&=
\frac{1}{2\rho_0}
\rho_{ij}
\biggl(\frac{m_i}{p_i^0}+\frac{m_j}{p_j^0}\biggr)
\sum_{k=1,2}
\biggl(\frac{1}{\mu_k{}^2}\biggr)
\frac{C_{k}}{(1-[p_{T,ij}/\mu_k]^2)^2}.
\end{align}

\begin{figure}[tbh]
\includegraphics[width=8.0cm]{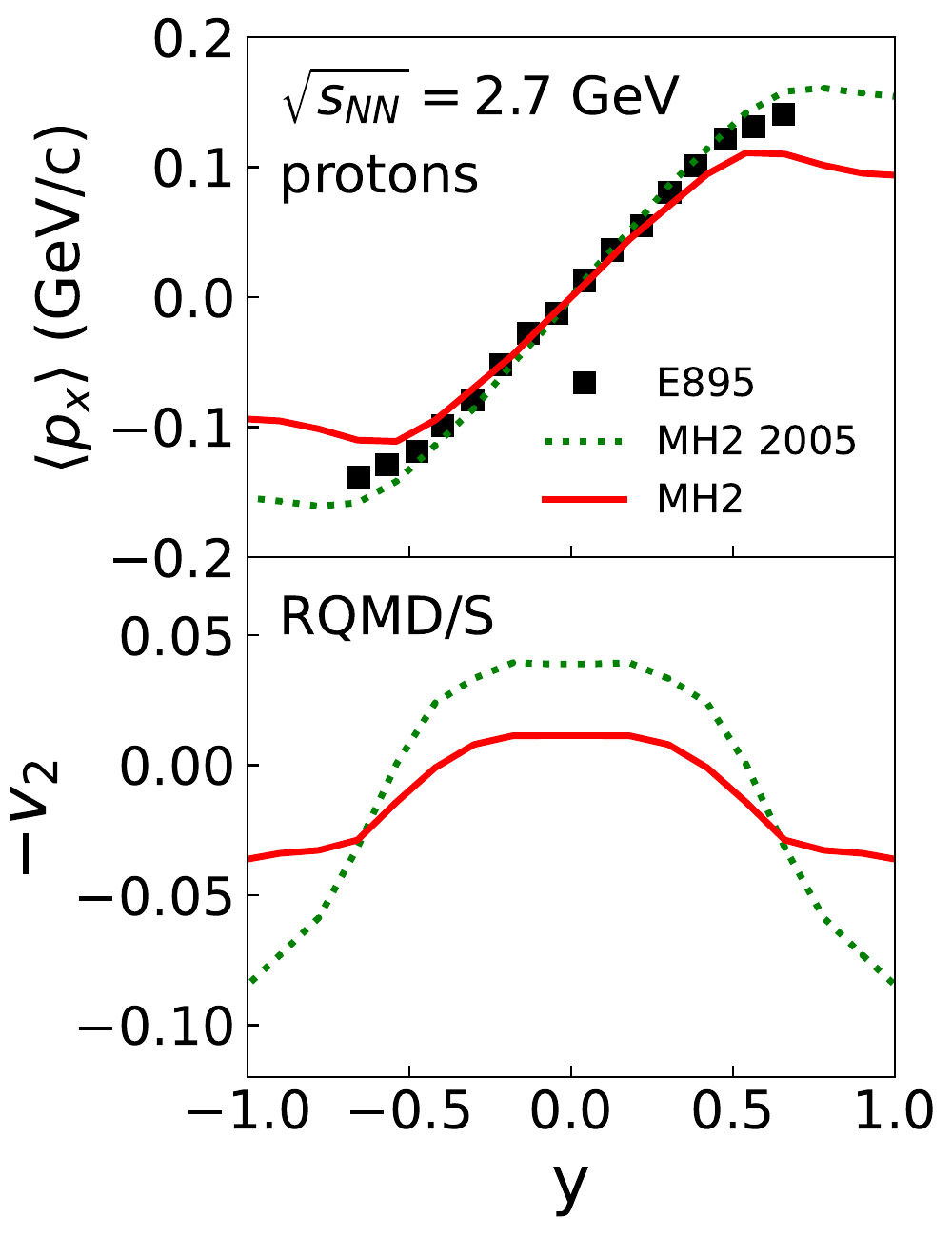}
\caption{The directed flow (upper panel)
and the elliptic flow (lower panel)
in Au + Au collisions at $\sqrt{s_{NN}}=2.7$ GeV.
Dotted lines correspond to the results of Ref.~\cite{Isse:2005nk},
in which the factor 2 larger force is used by mistake.
The corrected results are shown by the solid line.
the impact parameter range $4.0 < b < 8.0$ fm is used.
}
\label{fig:rqmds2005}
\end{figure}

The corrected results show less collective flow than the previous ones.
In the upper panel of Fig.~\ref{fig:rqmds2005}, 
the sideward flow in mid-central Au + Au collisions
at $\sqrt{s_{NN}}=2.7$ GeV is shown from RQMD/S.
The dotted line corresponds to the result,
in which the factor two is multiplied in the Skyrme force
to show the wrong result in Ref.~\cite{Isse:2005nk}.

\section{EoS parameters}
\label{appendix:eos}

The parameters of the potentials are fixed by the five conditions
assuming a nuclear matter binding energy $B=-16$ MeV
at a nuclear matter saturation density $\rho_0=0.168$ fm$^{-3}$
and the momentum-dependent parameter sets fulfill the conditions
$U_\mathrm{opt}(\rho_0,p=1700 \text{MeV})=60$ MeV and
$U_\mathrm{opt}(\rho_0,p=650\,\mathrm{MeV})=0$ MeV.
At the saturation density and $T=0$, we have the Weisskopf relation
\begin{equation}
\sqrt{m_N^2+p_f^2} + U_\mathrm{sk}(\rho_0) + U_m(p_f) = m_N + B
     \label{eq:e}
\end{equation}
where the Fermi momentum is $p_f=(6\pi^2\rho_0/g_N)^{1/3}$
with $g_N=4$ being the spin-isospin degeneracy.
Pressure $P=\rho^2\partial (e/\rho)/\partial\rho|_{\rho=\rho_0}$ is zero at the saturation density.
At zero temperature, the distribution function takes the form:
\begin{equation}
 f(x,p)=\frac{g_N}{(2\pi)^3}\theta(p_f-p)
    = g\theta(p_f-p),
\end{equation}
and pressure at zero temperature is given by
\begin{equation}
P = P_k + 
\frac{g}{2}\int^{p_f}_0 d^3p U_m(p)
 + \rho U_\mathrm{sk}(\rho)
 -\int^{\rho}_0 U_\mathrm{sk}(\rho')d\rho'
\end{equation}
where
\begin{align}
P_k 
  &=g \int_0^{p_f} d^3p \left(
   \frac{p^2}{3E}+\frac{p}{3}\frac{\partial U_m(p)}{\partial p}
     \right) \nonumber\\
  &= g \int_0^{p_f} d^3p \left(
   \frac{p^2}{3E} - U_m(p) 
  \right) + \rho U_m(p_f)
     \label{eq:p}
\end{align}
The nuclear incompressibility $K$ is defined by the second derivative
of the energy density with respect to the baryon density:
\begin{equation}
K=\left. 9\rho^2\frac{\partial^2}{\partial\rho^2}\left(
    \frac{e}{\rho}
   \right)\right|_{\rho=\rho_0}
  =\left.9\rho\frac{\partial^2e}{\partial\rho^2}\right|_{\rho=\rho_0}
\end{equation}
The first derivative of $e$ with respective to the baryon density
gives the baryon chemical potential:
\begin{equation}
 \mu = \frac{\partial e}{\partial\rho}=\sqrt{m^2+p_f^2}+U_\mathrm{sk}(\rho)+U_m(p_f)
\end{equation}
The incompressibility is now given by
\begin{equation}
 K=9\rho\frac{\partial\mu}{\partial\rho}
 =9\left( \frac{p_f^2}{3e} 
     + \rho\frac{\partial U_\mathrm{sk}}{\partial\rho}
     + \frac{p_f}{3}\frac{\partial U_m(p_f)}{\partial p_f}
     \right)
     \label{eq:k}
\end{equation}
where we used $\frac{\partial\rho}{\partial p_f}=3\rho/p_f$.
From Eqs.~(\ref{eq:e}), (\ref{eq:p}), and (\ref{eq:k}),
for the Skyrme potential Eq.~(\ref{eq:skyrme}) and
the momentum-dependent potential Eq.~(\ref{eq:momdep}),
we obtain the system of equations:
\begin{align}
& E_f + \alpha + \beta + U_m(p_f) = m_N + B,\\
&P_\mathrm{kin} + \frac{\alpha}{2}\rho_0
   + \frac{\beta\gamma\rho_0}{\gamma+1} 
   + \rho_0 U_m(p_f) -
\frac{g}{2}\int^{p_f}_0 d^3p U_m(p)
   = 0,\\
&\frac{p_f^2}{3e}+\alpha+\beta\gamma
 + \frac{p_f}{3}\frac{\partial U_m(p_f)}{\partial p_f}
  = \frac{K}{9} ,\\\
& \alpha + \beta  = U_\mathrm{opt}(\rho_0,p\to\infty),\\
& \alpha + \beta + U_m(p=650\,\mathrm{MeV}) = 0,
\end{align}
where $E_f=\sqrt{m^2+p_f^2}$ and
\begin{align}
 P_\mathrm{kin} &=d \int^{p_f}_0 d^3p\frac{p^2}{3\sqrt{m^2+p^2}} 
 = \frac{g_N}{16\pi^2}
 \nonumber\\
& \times\left[ \frac{2}{3}E_fp_f^3 - m^2E_fp_f
  + m^4\ln\left(\frac{E_f+p_f}{m}\right)
   \right]
\end{align}
We solve these equations for $\alpha,\beta,\gamma,\mu$, and $C$.

\section{scalar implementation}
\label{appendix:eos_scalar}


In the case of the scalar potential, we have
\begin{equation}
 \mu = \frac{\partial e}{\partial\rho}=\sqrt{m^{*2}(p_f)+p_f^2},
\end{equation}
and the incompressibility is given by
\begin{equation}
 K=9\rho\frac{\partial\mu}{\partial\rho}
 =9\left( \frac{p_f^2}{3e^*} 
     + \rho\frac{m^*}{e^*}\frac{\partial m^*}{\partial\rho}
     \right)
     \label{eq:ks}
\end{equation}
The derivative of the effective mass with respect to the density
can be calculated as
\begin{equation}
\frac{\partial m^*}{\partial\rho}=
 \frac{\partial U_s}{\partial\rho_s}\frac{\partial \rho_s}{\partial\rho}
 =\frac{\partial U_s}{\partial\rho_s}
  \left[
  \frac{m^*}{e^*}+ \frac{\partial m^*}{\partial\rho}
    \frac{g_N}{(2\pi)^3}\int d^3p\frac{p^2}{e^{*3}}
  \right]
\end{equation}
which yields~\cite{Matsui:1981ag}
\begin{equation}
\frac{\partial m^*}{\partial\rho}= \frac{\partial U_s}{\partial\rho_s}
  \frac{m^*}{e^*}
  \left[
  1 -  \frac{\partial U_s}{\partial\rho_s}
      \frac{g_N}{(2\pi)^3}\int d^3p\frac{p^2}{e^{*3}}
  \right]^{-1}
\end{equation}

\section{Derivatives of transverse distances}
\label{appendix:derivative}

The calculation of the derivatives for the transverse distances can be done as follows.
First, we consider the distance in the two-body c.m.:
\begin{eqnarray}
 q_{Tij}^2 &=& q_{ij}^2 - \frac{(q_{ij}\cdot P_{ij})^2}{s},\\
 p_{Tij}^2 &=& p_{ij}^2 - \frac{(p_i^2-p_j^2)^2}{s}
\end{eqnarray}
where $q_{ij}=q_i - q_j$, $p_{ij}=p_i- p_j$, $P_{ij}=p_i+p_j$, and $s=P_{ij}^2$.
The derivatives are
\begin{eqnarray}
\frac{\partial q_{Tij}^2}{\partial \bm{r}_{i}} &=&
  -2\left [\bm{r}_{ij} - \frac{(q_{ij}\cdot P_{ij})}{s}\bm{P}_{ij} \right]\\
\frac{\partial q_{Tij}^2}{\partial \bm{p}_i} &=&
   \frac{2(q_{ij}\cdot P_{ij})}{s}\left[
   \bm{r}_{ij} - P_{ij}^0 \frac{(q_{ij}\cdot P_{ij})}{s}
     \tilde{\bm{v}}_{ij}
     \right] \\
\frac{\partial p_{Tij}^2}{\partial \bm{p}_i} &=&
  -2\Big[
   \bm{p}_{ij} - 2(p_i^0- p_j^0)\frac{\bm{p}_i}{p^0_i}\\
    &&+ P_{ij}^0 \frac{(p_i^2-p_j^2)^2}{s^2}
     \tilde{\bm{v}}_{ij}
   \Big]
\end{eqnarray}
where $\tilde{\bm{v}}_{ij} =\bm{P}_{ij}/P_{ij}^0 - \bm{p}_i/p_i^0$.

The two-body distances in the rest frame of a particle $j$ are
\begin{eqnarray}
 q_{Rij}^2 &=& q_{ij}^2 - (q_{ij}\cdot u_{j})^2,\\
 p_{Rij}^2 &=& p_{ij}^2 - (p_{ij}\cdot u_j)^2,
\end{eqnarray}
where $u_j= p_j/m_j$.
The derivatives are 
\begin{eqnarray}
\frac{\partial q_{Rij}^2}{\partial \bm{r}_{i}} &=&
 -2\bm{r}_{ij} + 2(q_{ij}\cdot u_j)\bm{u}_j,\\
\frac{\partial q_{Rij}^2}{\partial \bm{p}_i} &=& 0 \\
\frac{\partial q_{Tji}^2}{\partial \bm{p}_i} &=& 
 -2\frac{(q_{ji}\cdot u_i)}{m_i}\bm{r}_{ji}
 \\
\frac{\partial p_{Rij}^2}{\partial \bm{p}_i} &=&
2\left[ 
p^0_{ij}\bm{v}_i
  -\bm{p}_{ij} 
   -\frac{(p_{ij}u_j)}{m_j} p_j^0\bm{v}_{ij}
     \right] \\
\frac{\partial p_{Rji}^2}{\partial \bm{p}_i} &=&
2\left[ 
p^0_{ij}\bm{v}_i
  -\bm{p}_{ij} 
   +\frac{(p_{ij}u_i)}{m_i} p_j^0\bm{v}_{ij}
     \right] \\
\end{eqnarray}
where $\bm{v}_{i}=\bm{p}_i/p_i^0$,
and $\bm{v}_{ij}=\bm{v}_i - \bm{v}_j$.

\section{Equations of motion}
\label{appendix:eom_rqmd}

\subsection{Equations of motion for RQMDs}

The equations of motion for the RQMDs model are
\begin{align}
\bm{\dot{x}_i} & =
   \frac{\bm{p}_i^{*}}{p_i^{*0}}
      +\sum_{j}
      \frac{m_j^*}{p_j^{*0}}
      \frac{\partial m_j^*}{\partial\bm{p}_i}
   \label{eq:EOMr}  ,\\
\bm{\dot{p}}_i
  &= -\sum_{j}
   \frac{m_j^*}{p_j^{*0}}
   \frac{\partial m_j^*}{\partial\bm{r}_i}
  \label{eq:EOMp}.
\end{align}
where $m_i^* = m_i + S_i$ and $S_i$ is the scalar potential:
\begin{equation}
 S_i  = V_i(\rho_{si}) + V_{m,i}(p_{Tij}^2)
\end{equation}
where the scalar density is given by
\begin{equation}
 \rho_{si}=\sum_{i(\neq j}f_j \rho_{ij},~~~f_j = m_j^*/p_j^{*0}
\end{equation}
and the density dependent part is
\begin{equation}
 V_i(\rho_{si}) = 
 \frac{\alpha}{2\rho_0} \rho_{si} + \frac{\beta}{(1+\gamma)}
  \left(\frac{\rho_{si}}{\rho_0}\right)^\gamma
\end{equation}
The momentum-dependent potential is given by Eq.~(\ref{eq:MVpotS}).
The equations of motion (\ref{eq:EOMr}) and (\ref{eq:EOMp}) become
\begin{eqnarray}
\bm{\dot{x}_i}  &=& 
   \frac{\bm{p}_i^{*}}{p_i^{*0}}
+ \sum_{j(\neq i)} \left[
   D_{ij}\frac{\partial q_{Tij}^2}{\partial \bm{p}_i}
  +D_{ji}\frac{\partial q_{Tji}^2}{\partial \bm{p}_i}
  \right]\\
  &+& \sum_{j(\neq i)} \left[
   E_{ij}\frac{\partial p_{Tij}^2}{\partial \bm{p}_i}
  +E_{ji}\frac{\partial p_{Tji}^2}{\partial \bm{p}_i}
  \right],\label{eq:rqmdsr}\\
\bm{\dot{p}_i}  &=& -\sum_{j(\neq i)} \left[
   D_{ij}\frac{\partial q_{Tij}^2}{\partial \bm{r}_i}
  +D_{ji}\frac{\partial q_{Tji}^2}{\partial \bm{r}_i}
  \right]. \label{eq:rqmdsp}
\end{eqnarray}
where 
\begin{eqnarray}
D_{ij} &= &\frac{m_i^*}{p_i^{*0}}f_j \left(
    \frac{\partial V_i(\rho_{si})}{\partial \rho_{si}}
    + V_{m,ij}
    \right)
    \frac{\rho_{ij}}{4L},\\
E_{ij} &= &\frac{m_i^*}{p_i^{*0}}f_j\frac{\partial V_{m,ij}}{\partial p_{Tij}^2}\rho_{ij},\\
V_{m,ij} &=& \frac{C}{2\rho_0} \frac{1}{1-p_{Tij}^2/\mu^2}.
\end{eqnarray}
If $q_{Tij}$ is defined as the distance in the two-body c.m.,
$q^2_{Tij}=q^2_{Tji}$,
one may need to add the term that comes from the derivative of the factor
$f_i$
in the scalar density
\begin{equation}
 \sum_{j(\neq i)} \frac{m_i^*}{p_i^{*0}}f_j\frac{\partial S_i}{\partial\rho_{si}}\rho_{ij}
\frac{\partial f_i}{\partial \bm{p}_i}
\end{equation}

\subsection{Equations of motion for RQMDv}

The equations of motion for RQMv are
\begin{align}
\bm{\dot{x}_i} & =
   \frac{\bm{p}_i^{*}}{p_i^{*0}}
    +\sum_{j}
    v^{*\mu}_j
    \frac{\partial {V}_{j\mu}}{\partial\bm{p}_i}
    , \label{eq:EOMvr}\\
\bm{\dot{p}}_i
   &= -\sum_{j}
   v^{*\mu}_j
    \frac{\partial V_{j\mu}}{\partial\bm{r}_i}
    \label{eq:EOMvp}.
\end{align}

The equations of motion (\ref{eq:EOMvr}) and (\ref{eq:EOMvp})
for the vector implementation have the same structure as
(\ref{eq:rqmdsr}) and (\ref{eq:rqmdsp}),
but different $D_{ij}$ and $E_{ij}$:
\begin{eqnarray}
D_{ij} &=& 
\frac{\rho_{ij}}{4L}
v^{*\mu}_i 
(B_i B_j A_{ij\mu} +  u^*_{j\mu}V_{m,ij}) 
,\\
A_{ij\mu} &=& \frac{J_i \cdot u^*_j}{\rho_{Bi}}
  \frac{\partial}{\partial\rho_{Bi}}\left(\frac{V_i}{\rho_{Bi}}\right)
   J_{i\mu} + \frac{V_i}{\rho_{Bi}}u^*_{j\mu},\\
E_{ij} &=& v_i^{*\mu}u^*_{j\mu} \frac{\partial V_{m,ij}}{\partial
p_{Tij}^2}\rho_{ij},\\
  J^\mu_i &=& \sum_{i(\neq j)}B_j u_j^{*\mu} \rho_{ij},
\end{eqnarray}
where $B_i$ is the baryon number and
$V_i$ is a function of an invariant baryon density $\rho_{Bi}=\sqrt{J_i^2}$.

\end{document}